\documentclass[12pt]{article}

\usepackage{amsmath,amsthm, amsfonts, amssymb, amsxtra, amsopn}
\usepackage{pgfplots}
\pgfplotsset{compat=1.13}
\usepackage{pgfplotstable}
\usepackage{graphicx,grffile}
\usepackage{multirow}
\usepackage{booktabs}
\usepackage{listings}
\usepackage{cmap}
\usepackage{colortbl}
\usepackage{adjustbox}
\usepackage{epsfig}

\usepackage[tableposition=top]{caption}
\usepackage{subcaption}
\usepackage{makecell} 

\PassOptionsToPackage{hyphens}{url}

\usepackage{hyperref}
\hypersetup{colorlinks=true,linkcolor=black,citecolor=black,urlcolor=blue,filecolor=black}
\hypersetup{pdfpagemode=UseNone,pdfstartview=}

\usepackage{enumitem}
\setlist[itemize]{noitemsep, topsep=0pt}

\advance\oddsidemargin by -0.2in
\advance\textwidth by 0.4in

\advance\topmargin by -0.25in
\advance\textheight by 0.5in

\long\def\symbolfootnotetext[#1]#2{\begingroup%
\def\thefootnote{\fnsymbol{footnote}}\footnotetext[#1]{#2}\endgroup}

\DeclareMathOperator{\TP}{TP}
\DeclareMathOperator{\FP}{FP}
\DeclareMathOperator{\TN}{TN}
\DeclareMathOperator{\FN}{FN}

\def\zz{\phantom{0}}

\title{Generative Adversarial Networks and Image-Based Malware Classification}

\author{Huy Nguyen\footnotemark[1]\ \ \ 
Fabio Di Troia\footnotemark[1]\ \ \
Genya Ishigaki\footnotemark[1]\ \ \
Mark Stamp\footnotemark[1]\,\,\footnotemark[2]}

\begin{document}

\symbolfootnotetext[1]{Department of Computer Science, San Jose State University}
\symbolfootnotetext[2]{mark.stamp$@$sjsu.edu}

\maketitle

\abstract
For efficient malware removal, determination of
malware threat levels, and damage estimation, malware family classification plays 
a critical role. 
In this paper, we extract 
features from malware executable files and represent them as images using 
various approaches. We then focus on Generative Adversarial Networks (GAN) 
for multiclass classification and compare our GAN results to other popular 
machine learning techniques, including Support Vector Machine (SVM), 
XGBoost, and Restricted Boltzmann Machines (RBM).
We find that the AC-GAN discriminator is generally
competitive with other machine learning techniques.
We also evaluate the utility of the GAN generative model for
adversarial attacks on image-based malware detection. 
While AC-GAN generated images are visually impressive, we find that they
are easily distinguished from real malware images using any of several learning
techniques. This result indicates that our GAN generated images would be
of little value in adversarial attacks.

\section{Introduction}

The Covid-19 pandemic, which has run amok worldwide for more than two years, 
has drastically increased the trend of working from home.  
The remote work environment has also pushed another trend: 
increasing cyber attacks, including phishing, data breaches, and malware. 
According to CSO~\cite{carlson2021},
in the second quarter of 2020, as compared to the same period a year earlier,
cloud security incidents increased by~188\%, ransomware attacks 
grew by over~40\%, and email malware attacks were up by~600\%.

Malware, short for ``malicious software'', consists of computer programs that are written 
to cause harm to computer and Internet users~\cite{RAZAK201658}.
Common types of malware include botnets, rootkits, Trojans, worms, 
and spyware. Malware can be used to steal information, utilize hardware, 
cause disruption for financial or reputational gain, or other unauthorized activity. 
Malware defense is an ongoing battle with multiple layers: preventing malware from entering, 
alerting users that a system is compromised, removal of malware
from compromised systems, and so on.

There has been considerable research into 
malware detection and classification. In recent years,
malware classification based on 
machine learning has become a leading focus of such research. 
With more computing power from graphic processing units (GPU) and Google 
tensor processing units (TPU), which are specialized for machine learning techniques
and feature extraction, costly deep learning image-based malware detection 
techniques have become a viable option. 

An image can be derived directly from the byte 
sequence of an executable file without executing (or emulating)
or otherwise pre-processing the data to extract features. Powerful image-based 
techniques, including Convolutional Neural Networks (CNN) and Generative 
Adversarial Networks (GAN) have been used to classify malware samples
with impressive results.

Malware detection is an arms race between detectors and malware writers, 
where each side tries to develop new and innovative ways to defeat the other side. 
Adversarial attacks on machine learning based malware defenses have
been developed.
In one type of adversarial attack, a malware writer attempts to contaminate
the training data, so that the resulting model is less effective. A possible
approach to such an adversarial attack on an image-based malware 
detection system is to generate ``deep fake'' 
malware images to pollute the training dataset. 

GANs have been used to generate realistic fake images.
Researchers at Nvidia~\cite{Karras} developed StyleGAN, 
a style-based architecture for GAN, which by some measures was~20\%\ more
effective than a traditional GAN generator. StyleGAN was 
also used to create the trending website ``thispersondoesnotexist.com''.
We consider the utility of GANs for adversarial attacks
on image-based malware systems. 
MalGAN is a GAN technique that was designed specifically 
to deal with malware images~\cite{malgan,improved_malgan}.

In addition to malware detection, malware classification is critically important, as it enables us 
to estimate the damage, determine the threat level, and to provide protection 
specific to a given malware family. In this research, we employ auxiliary classifier GAN (AC-GAN) 
for multi-class classification of malware families and compare with a wide variety of other 
machine learning models, including 
Support Vector Machine (SVM)~\cite{svm}, 
$k$-Nearest Neighbors ($k$-NN)~\cite{knn},
multilayer perceptron (MLP)~\cite{StampBook}, 
Random Forest (RF)~\cite{randomforest}, 
Restricted Boltzmann Machines (RBM)~\cite{rbm2012}, 
XGBoost~\cite{xgboost}, 
as well as the deep residual network,
Resnet152~\cite{resnet}. 
We also develop SVM models to test the quality of fake images generated by AC-GAN generative model.
The dataset that we use consists of more than~26,000 malware executables from~20 
distinct families~\cite{dataset}. 

Among the contributions of this paper are the following.
\begin{itemize}
\item We compare four distinct techniques for generating images from malware samples.
Both grayscale and color images are considered. As far as the authors are aware,
no previous work has provided a direct comparison of all of the image generation techniques
considered in this paper.
\item We apply a wide variety of learning techniques to a challenging malware image classification problem.
In each case, we carefully tune the hyperparameters, which results in a meaningful comparison
of these learning techniques in the image-based malware classification domain. 
The only comparable work that the authors are aware of is~\cite{Pratik}, which is focused
on sequential techniques, including Long Short-Term Memory (LSTM) and 
Gated Recurrent Unit (GRU) classifiers.
\item We carefully consider the potential utility of GAN-generated
``deep fake'' malware images for adversarial attacks.
While such deep fake images are indeed visually impressive, we find that they
are woefully inadequate for adversarial attacks. The authors are not aware of any
previous work that explicitly makes this point, with 
the possible exception of~\cite{nagaraju2021auxiliaryclassifier},
which is primarily focused on the classification potential of GANs,
rather than their potential in adversarial attacks.
\end{itemize}

The remainder of this paper is organized as follows.
In Section~\ref{chap:background}, we discuss related work and briefly
consider each of the 
various machine learning and deep learning models 
used in our research. 
Our dataset and image extraction techniques are introduced
in Section~\ref{sect:dataset}.
In Section~\ref{chap:result}, we present our extensive experimental results, 
and we provide analysis and discussion of these results.
Section~\ref{chap:future} concludes the paper, and we outline 
a few potential avenues for future work.

\section{Background}\label{chap:background}

New malware families are being developed everyday. For example, 
it was not too long ago that the SolarWinds zero day attack caused damage 
to several government agencies~\cite{LAZAROVITZ202117}. The attack was caused 
by a state-sponsored group and put the entire cyber security industry on high 
alert. The SolarWinds attack reminded us of the importance of malware 
defense, and malware detection in particular. 

There are two broad categories of malware detection: signature-based and anomaly-based. 
Signature-based malware detection keeps certain characteristics of 
previously-seen malware in a dictionary and detects based on this previously-determined information. 
There are three main disadvantages of a signature-based approach: the size of the dictionary may 
not be scalable, new malware or zero day attacks cannot be detected, and signatures can 
be changed by malware writers using obfuscation techniques. Anomaly-based detection 
tries to distinguish between benign and malware based on general characteristics or behavior. 
One disadvantage of anomaly-based detection is that the training dataset can be 
polluted, making anomalous features appear to be normal. 

Machine learning techniques can be viewed as a type of anomaly-based detection,
but such techniques also can be considered as higher-level ``signatures'', in the sense that a model
captures characteristics of an entire family, rather than a single instance. Regardless of how
we view machine learning models, adversarial attacks are a major concern.
One purpose of our research is to consider the effectiveness of generated
fake malware images in such adversarial attacks---with the ultimate goal of
more robust learning models.

Next, we discuss relevant related work. Then we 
briefly introduce each of the learning techniques considered in this paper.
We conclude this section with a discussion of the statistics used
to quantify classification success in our experiments.

\subsection{Related Work}

As a first example of research in image-based analysis, Jain and 
Stamp~\cite{jain2019image} 
used Convolutional Neural Networks (CNN) and Extreme Learning Machines (ELM) 
to classify malware. 
Jain and Stamp suggested future work including
testing different techniques for image extraction, 
such as zero padding and GIST 
descriptors of images.

In another paper, Nagaraju and Stamp~\cite{nagaraju2021auxiliaryclassifier} 
worked on image-based malware analysis using the GAN architecture known as 
Auxiliary-Classifier GAN(AC-GAN). They experimented with different image sizes 
from $32\times 32$, $64\times 64$ to as large as $512\times 512$. Grayscale images were extracted 
and truncated from executable files to the desired sizes. In addition to AC-GAN, 
CNN and ELM were considered, with CNN achieving impressive 
results in detecting fake images. For future work, they advised researchers 
to consider cutting-edge models, including VG-199 and ResNet152.

Xiao et al.~\cite{XIAO2021102420} introduced a novel framework called MalCVS,
which is their shorthand for Malware 
Classification using Colab image generation, VGG16, and SVM. 
Images were generated similar to grayscale, but with thick colored lines between
each section in the executable files. The images then were passed 
through VGG16 for feature extraction with the results fed to a multi-class SVM for 
classification. The MalCVS framework achieved impressive results with 
98.94\% accuracy and F1-score at 97.91\% on a multiclass (16 family) problem. 
Note that the MalCVS framework 
depart from strict grayscale images by its use of color borders in its image 
representation of malware.

A recent trend in image-based malware analysis is the use of color images.
Both Vasan et al.~\cite{VASAN2020107138} 
and Singh et al.~\cite{SINGH2019} generate color maps and use byte sequences from 
the executable files to represent color images. Both papers also use similar 
learning techniques, including CNN and Residual Neural Networks (specifically, ResNet-50). 
Singh et al. achieved impressive results with the MalImg dataset, obtaining an accuracy of 
98.10\%\ using ResNet-50, while Vasan et al.
attained an accuracy of 98.82\%\ using a so-called
``fine-tuned'' CNN architecture.
These papers tend to indicate that color (RGB) representations captured more 
pattern information, and therefore can achieve better results 
as compared to grayscale malware images.

The idea of generating fake features for obfuscation and adversarial attacks 
on malware systems is not new. For example, Hu et al.~\cite{malgan} 
and Kawai et al.~\cite{improved_malgan} proposed MalGAN to bypass black-box machine learning based 
detection models. MalGAN uses the output of a black-box model and employs GAN 
to generate fake samples.
MalGAN has been shown to be capable of substantially decreasing the 
true positive rate (TPR) to near~0.
The process of training MalGAN is fast and efficient, making it 
practical for such attacks.

In the area of multiclass classification, Fu et al.~\cite{section_colored} 
achieved impressive results with a accuracy of~97.47\% and F-measure 
of~96.85\% in categorizing~15 different malware families. They focused heavily on 
extracting features from malware executables, with a combination of global and local features. 
Color features were extracted using a Portable Executable (PE) format parser, 
and color images were built using RGB layers. The executable files were divided into sections,
with features such as entropy, byte sequences, relative size, and so on, extracted per section.
The results were then combined to represent the different RGB channels. The resulting images were 
colorful and enable us to visualize the data and code sections in malware files. 
For future work, they suggested that deep learning models such as CNN could be developed 
for malware classification. 

Farhat and Rammouz~\cite{resnet152} experimented with several pre-trained 
deep convolutional models, including VGG16, 
Resnet50, Resnet152, and MobileNet. 
With MobileMet, for example, they obtained an accuracy of~94\% after just 
one training epoch for a 9-class malware classification problem. 
After 25 epochs, the accuracy reached a peak of more than~97\%.
Pre-trained models take advantage of powerful image-based techniques
and have been trained on vast datasets, while only requiring fine tuning for
the specific task at hand. Thus, we can use 
pre-trained models for computer vision tasks and save a tremendous 
amount of training time.

In our research, we consider a wide variety of machine learning techniques,
specifically, SVM, $k$-NN, MLP, RBM, and two ensemble techniques:, namely, 
Random Forest and XGBoost. We also consider three deep learning models: 
DC-GAN, AC-GAN and Resnet152. In the next section, we briefly discuss each 
of these techniques, and we mention the relative advantages and 
disadvantages of each in the context of image-based malware classification. 

\subsection{Machine Learning Models}

As its name suggests,
the $k$-Nearest Neighbors ($k$-NN) classifier simply uses 
the~$k$ nearest neighbors from the training 
set to classify a new sample. Consequently, $k$-NN is simple and
easy to understand, and it perform surprisingly well in many applications. 
However, with images of size~$128\times 128 \times 3$, for example, 
the feature vector would be of length~49152, making distance computations 
in $k$-NN slow.

\begin{figure}[!htb]
    \centering
    \includegraphics[width=0.5\textwidth]{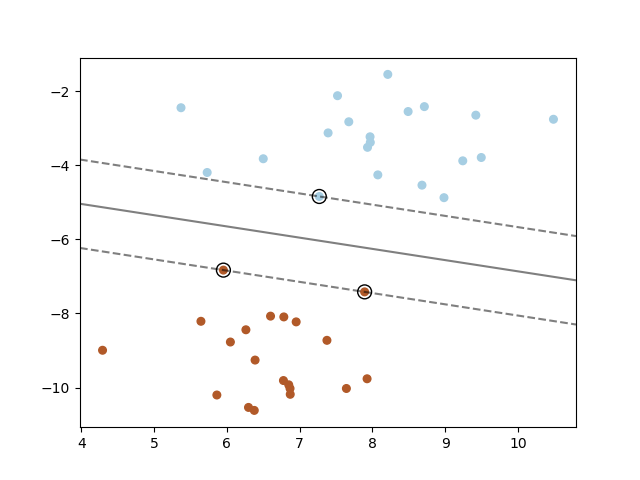}
    \caption{SVM separating hyperplane~\cite{svm_sklearn}} 
    \label{fig:svm_hyperplane}
 \end{figure}

Support Vector Machine (SVM) and the multiclass version, Support Vector Classifier (SVC), 
use a hyperplane to separate and classify data. Figure~\ref{fig:svm_hyperplane} illustrates a 2D 
version of SVM hyperplane---in this case, a line separating two classes of data. 
SVM performs well on many malware problems; for example, SVM
achieved 93.20\%\ accuracy on a 25-class classification problem in~\cite{VASAN2020107138}.
However, similar to $k$-NN, when the feature space is large, SVM is quite slow. 

We also consider
two ensembles: Random Forest (RF) and \hbox{XGBoost}. An ensemble is a group of models that 
are combined in some way to function as a whole. 
XGBoost requires substantial memory, as the dataset needs to be preprocessed before training.  

 \subsection{Deep Learning Models}

Generative Adversarial Networks (GAN) are innovative techniques
where a generative and discriminative model are trained simultaneously. 
The two models compete with each other, and thus can yield improved 
results, as compared to training each individually. 
There are many GAN variants, including ProGAN, StyleGAN, and so on. 
We experiment with DC-GAN for unsupervised generative models 
and AC-GAN for multi-class discrimination. 
A primary goal for this research is to examine both the discriminator and the generator 
models of GANs and compare them with various machine learning techniques. 
The discriminator can be used for a multiclass classification of malware families 
and binary classification of fake vs real images. The generator of different GAN 
architectures are compared against each other and against other machine learning
techniques.

Additionally, we consider Restricted Boltzmann Machines (RBM)
and the deep Residual Network (ResNet) 
architecture known as Resnet152. 
The Resnet152 model has been pre-trained on a vast image dataset. In
previous work, this particular model has been shown to achieve state-of-the-art results for
malware problems. Next, we briefly introduce each of these deep learning models.

\subsubsection{Deep Convolutional GAN (DC-GAN)}

DC-GAN was first introduced by Radford and Metz~\cite{radford2015unsupervised} 
in 2015 as an unsupervised technique for representation learning. 
As an unsupervised architecture, DC-GAN is applicable to 
unlabelled data. The resulting models from DC-GAN can be used to compare 
performance with other GAN architectures in binary classification of real vs fake images. 

\begin{figure}[!htb]
\centering
\includegraphics[width=0.8\textwidth]{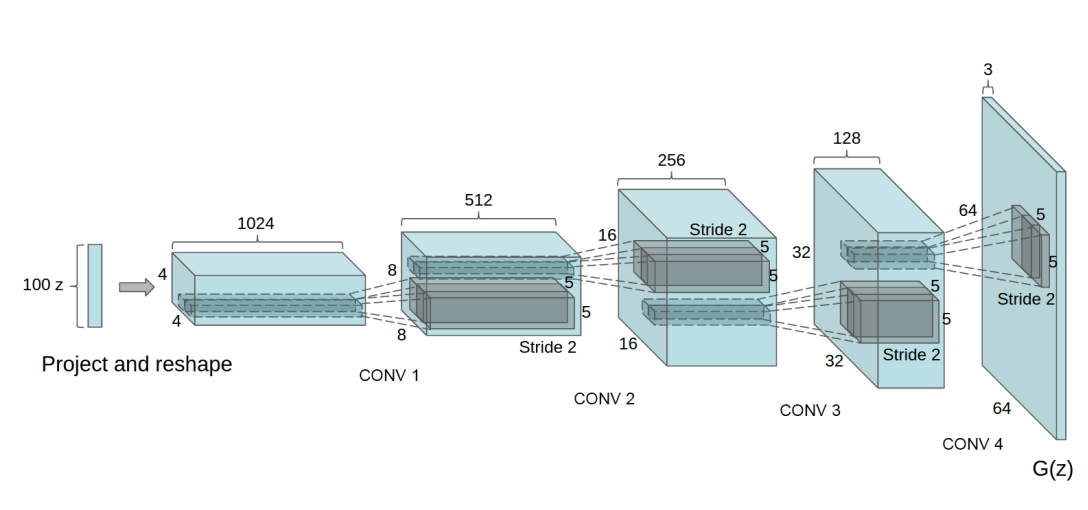}
\vglue-0.2in
\caption{DC-GAN generator \cite{radford2015unsupervised}} 
\label{fig:DCGAN_architecture}
\end{figure}

Figure~\ref{fig:DCGAN_architecture} illustrates the convolutional layers of 
DC-GAN without fully connected or pooling layers. A noise vector consists 
of~100 random numbers is fed into the model and generates images based on 
training data. The basic architecture of DCGAN has four convolutional layers 
and the output is expected to be~$64\times 64\times 3$. We can modify the settings in 
the convolutional layers to work with our grayscale~$128\times 128\times 1$, 
$256\times 256\times 1$ or even RGB images~$128\times 128\times 3$.

\subsubsection{Auxiliary-Classifier GAN (AC-GAN)}

There are multiple research papers showing that AC-GAN performs well with multiclass 
data~\cite{kang2021rebooting,nagaraju2021auxiliaryclassifier,odena2017conditional}. 
In contrast to DC-GAN, AC-GAN makes use of class labels. Using this extra data (the class labels), 
can help us to generate samples from a specific family when desired. The discriminator 
can also be used to solve the multiclass classification problem, which is our 
main focus in this research. 

\begin{figure}[!htb]
    \centering
    \includegraphics[width=0.375\textwidth]{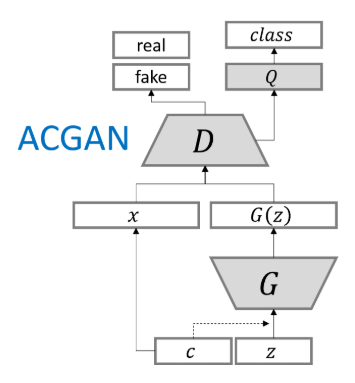}
    \caption{AC-GAN architecture \cite{acgan-figure1}} 
    \label{fig:acgan_architecture}
\end{figure}

As we can see in Figure~\ref{fig:acgan_architecture}, 
input~$C$ representing class labels is fed into both the generator and the discriminator. 
There are two outputs for the discriminator: validity of the image and the class label.
The discriminator is trained based on these two outputs, with two loss functions.

\subsubsection{Restricted Boltzmann Machines}

Restricted Boltzmann Machines (RBM) can extract non-linear features from images,
which can then be used by 
a linear model, such as Logistic Regression, In some applications, RBMs
perform well, for example, RBMs achieve 
94\% accuracy for a digit classification problem~\cite{rbm_sklearn}. Both RBMs 
and Logistic Regression are fast and efficient to train. 

\begin{figure}[!htb]
    \centering
    \vglue -4pt
    \includegraphics[width=0.65\textwidth]{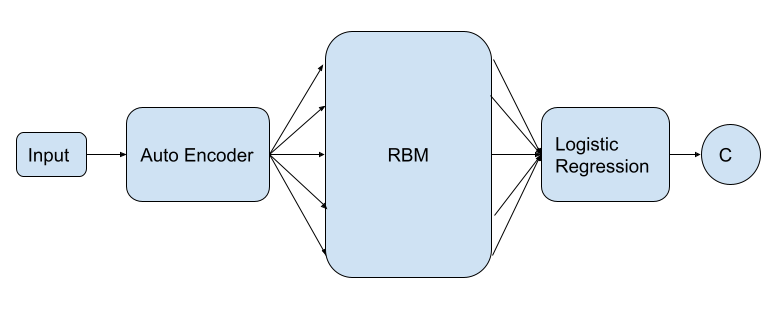}
    \vglue -10pt
    \caption{RBM architecture} 
    \label{fig:rbm_architecture}
\end{figure}

For our RBM models, we use BernoulliRBM, the implementation details 
for which can be found in~\cite{BernoulliRBM}.
The RBM acts as a layer that can be viewed as extracting meaningful 
smaller images from malware images. Then the resulting images are 
fed into a Logistic Regression layer for multiclass classification. 
We also add an AutoEncoder layer in front of the RBM to reduce noise in the images. 
Figure~\ref{fig:rbm_architecture} illustrates the RBM architecture from a high-level perspective.

\subsubsection{Resnet152}

\begin{figure}[!htb]
    \centering
    \vglue -4pt
    \includegraphics[width=0.225\textwidth]{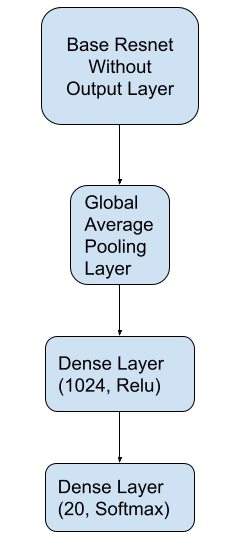}
    \vglue -2pt
    \caption{ResNet152 layers} 
    \label{fig:RN_layers}
\end{figure}

For ResNet152 models, 
we use the Tensorflow and Keras packages. Specifically, we use Resnet152v2, which 
has been pre-trained on the ImageNet dataset~\cite{imagenet}. 
To change the input size and class label, we add one dense layer after the base 
ResNet output, we freeze all the base ResNet model weights, and then train the extra layer. 
Finally, we unfreeze half of the layers of the pre-trained base models and train 
for several epochs. Figure~\ref{fig:RN_layers} provides a high-level 
illustration of the ResNet152 architecture used in our experiments
discussed in Section~\ref{chap:result}.

\subsection{Evaluation Metrics}

For the multiclass classification problem, to distinguish malware families 
from each other, we focus on the AC-GAN discriminator model. 
We then compare the performance of AC-GAN discriminator with other machine learning models 
including SVM, RBM, and XGBoost. For all of our experiments,
our evaluation metrics are accuracy, precision, recall, and f1-score, which are calculated 
based on the number of
True Positive (TP), False Positive (FP), True Negative (TN), 
False Negative (FN) results. The accuracy is computed as
$$
\mbox{accuracy} = \frac{\TP+\TN}{\TP+\TN+\FP+\FN}
$$
while precision and recall are given by
$$
\mbox{precision} = \frac{\TP}{\TP+\FP}\mbox{\ \ and\ \ }\mbox{recall} = \frac{\TP}{\TP+\FN}
$$
and, finally, the f1-score is computed as
$$
\mbox{f1} = \frac{2\times\mbox{Precision}\times
	\mbox{Recall}}{\mbox{Precision}+\mbox{Recall}} = \frac{2 \TP}{2 \TP+\FP+\FN}
$$



Receiver operating characteristic (ROC) curves provide a way to 
quantify the performance of a classifier. Specifically, the area under the ROC curve (AUC)
can measures the probability that a randomly selected positive instance
scores higher than a randomly selected negative instance~\cite{BRADLEY19971145}. 
For our fake malware image detection experiments, which are discussed in Section~\ref{sect:GIP},
we employ AUC as a metric.



\section{Data and Features}\label{sect:dataset}

In this section, we first discuss the dataset that we have used
in some detail. Then we introduce
the various image extraction techniques that we employ
to generate the features for our machine learning experiments.

\subsection{Dataset}

\begin{figure}[!htb]
\centering
\begin{tikzpicture}[scale=0.7, every node/.style={scale=1.0},rotate=-90]
    \begin{axis}[
        width  = 0.7*\textwidth,
        height = 14cm,
        ymin=0,ymax=4000,
        ytick={0,500,1000,1500,2000,2500,3000,3500},
        major x tick style = transparent,
        ybar=5*\pgflinewidth,
        bar width=6.5pt,
        xlabel = {Family},
        xlabel style={rotate=180},
        ylabel = {Number of samples},
        yticklabel pos=right,
        symbolic x coords={
    Adload,
    Agent,
    Alureon, 
    Bho, 
    Ceeinject, 
    Cycbot, 
    Delfinject, 
    Fakerean, 
    Hotbar, 
    Lolyda, 
    Obfuscator, 
    Onlinegames, 
    Rbot, 
    Renos, 
    Startpage, 
    Vobfus, 
    Vundo, 
    Winwebsec, 
    Zbot, 
    Zeroaccess
	},
        xtick={
    Adload, 
    Agent, 
    Alureon, 
    Bho, 
    Ceeinject, 
    Cycbot, 
    Delfinject, 
    Fakerean, 
    Hotbar, 
    Lolyda, 
    Obfuscator, 
    Onlinegames, 
    Rbot, 
    Renos, 
    Startpage, 
    Vobfus, 
    Vundo, 
    Winwebsec, 
    Zbot, 
    Zeroaccess
	},
	y tick label style={
		rotate=90,
    		/pgf/number format/.cd,
		1000 sep={},
   		fixed,
   		fixed zerofill,
    		precision=0},
        x tick label style={
        		rotate=90,
		font=\small\tt,
		},
        nodes near coords,
        every node near coord/.append style={rotate=90, 
        								   anchor=west,
								   font=\footnotesize,
								   /pgf/number format/.cd,
								   	fixed zerofill,
									1000 sep={},
									precision=0
								   },
        enlarge x limits=0.045,
    ]
\addplot[fill=blue,opacity=1.00] 
coordinates {
(Adload,1050) 
(Agent,842) 
(Alureon,1328) 
(Bho,1176) 
(Ceeinject,894) 
(Cycbot,1029) 
(Delfinject,1146) 
(Fakerean,1063) 
(Hotbar,1491) 
(Lolyda,915) 
(Obfuscator,1445) 
(Onlinegames,1293) 
(Rbot,1017) 
(Renos,1312) 
(Startpage,1136) 
(Vobfus,926) 
(Vundo,1793) 
(Winwebsec,3651) 
(Zbot,1786) 
(Zeroaccess,1119)
};
\end{axis}
\end{tikzpicture}
\caption{MalExe samples per family} 
\label{fig:malexe_size}
\end{figure}
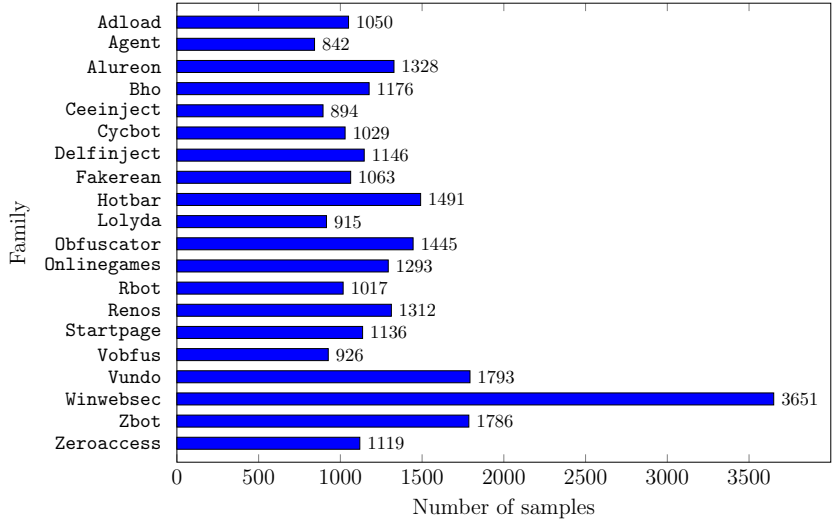

The malware data that we use in the experiments discussed in this paper 
is derived from the MalExe dataset~\cite{dataset,SKim}. The part of this vast 
dataset that we consider
consists of~26,412 malware executable files from~20 
different families. Each family has between~842 and~3651 samples. 
Figure~\ref{fig:malexe_size} shows the number of samples available in each family. 
Note that three families (Vundo, Winwebsec, and Zeroaccess) have the
most samples, while most of the other families have similar amounts of samples. 
Table~\ref{malware_overview} summarizes each the~20 malware families.

\begin{table}[!htb]
    \caption{Malware families\label{malware_overview}}
    \centering
    \adjustbox{scale=0.775}{
    \begin{tabular}{c|cc}\midrule\midrule
    Family & Type & Description \\ \midrule
    Adload & Adware & Shows ads, poses high threat \cite{adload} \\
    Agent & General & Performs malicious actions \cite{agent}  \\
    Alureon & Trojan & Steals information \cite{alureon} \\
    Bho & Trojan & Steals information, redirects web sites \cite{bho} \\ 
    Ceeinject & Virtool & Obfuscates itself to hide purposes \cite{ceeinject} \\
    Cycbot & Backdoor/Trojan & Provides backdoor access \cite{cycbot} \\
    Delfinject & PWS & Steals passwords \cite{delfinject}\\
    Fakerean & Rogue & Raises false alarms to make money \cite{fakerean} \\ 
    Hotbar & Adware & Displays advertisements \cite{hotbar} \\
    Lolyda & PWS & Monitors network activities \cite{lolyda}  \\
    Obfuscator & Virtool & Obfuscates itself to hide purposes \cite{obfuscator}  \\
    Onlinegames & PWS/Trojan & Injects malicious files, steals information \cite{onlinegames} \\ 
    Rbot & Backdoor/Trojan & Provides backdoor access\cite{rbot} \\
    Renos & Trojan & Downloads unwanted softwares \cite{renos}  \\
    Startpage & Trojan & Changes internet browser homepage \cite{startpage}  \\
    Vobfus & Worm & Downloads and spreads malwares \cite{vobfus} \\ 
    Vundo & Trojan Downloader & Advanced defensive and stealth techniques \cite{vundo} \\
    Winwebsec & Rogue & Raises false alarms for money \cite{winwebsec}  \\
    Zbot & Trojan & Steals information, gives access to hackers \cite{zbot}  \\
    Zeroaccess & Trojan & Disables security features \cite{zeroaccess} \\ 
    \midrule\midrule
    \end{tabular}
    }
    \end{table}


\begin{figure}[!htb]
\centering
\begin{tikzpicture}[scale=0.75, every node/.style={scale=1.0}]
    \begin{axis}[
        width  = 1.0*\textwidth,
        height = 8cm,
        xmin=0,xmax=1000,
        ymin=0,ymax=7250,
        ytick={0,1000,2000,3000,4000,5000,6000,7000},
        major x tick style = transparent,
        ybar=5*\pgflinewidth,
        bar width=35.0pt,
        xlabel = {Files size in KB},
        ylabel = {Number of samples},
        xtick={
0,
100,
200,
300,
400,
500,
600,
700,
800,
900,
1000
	},
	y tick label style={
		font=\small,
    		/pgf/number format/.cd,
   		fixed,
   		fixed zerofill,
		1000 sep={},
    		precision=0},
        x tick label style={
		font=\small,
    		/pgf/number format/.cd,
   		fixed,
   		fixed zerofill,
		1000 sep={},
    		precision=0
		},
        enlarge x limits=0.05,
    ]
\addplot[fill=blue,opacity=1.00] 
coordinates {
(50,6000)
(150,7000)
(250,3000)
(350,3250)
(450,3000)
(550,1500)
(650,1150)
(750,500)
(850,350)
(950,500)
};
\end{axis}
\end{tikzpicture}
\caption{Histogram of MalExe file sizes} 
\label{fig:size_histogram}
\end{figure}
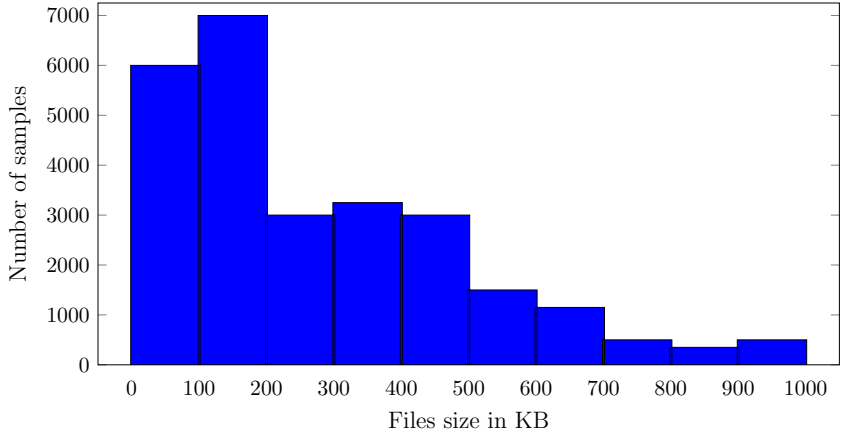

Figure~\ref{fig:size_histogram} 
shows the histogram of file sizes over the entire MalExe dataset. We see that 
the majority of samples are smaller than 200KB while we have a fair amount of 
samples between 200-500KB and a small number of samples larger than 500KB 
in size. Most image-based analysis choose an image size of $256\times 256$ or $224\times 224$, 
and without resizing, that would equate to 64KB and 49KB, respectively. We note that most of our 
samples have enough data to form such images. 


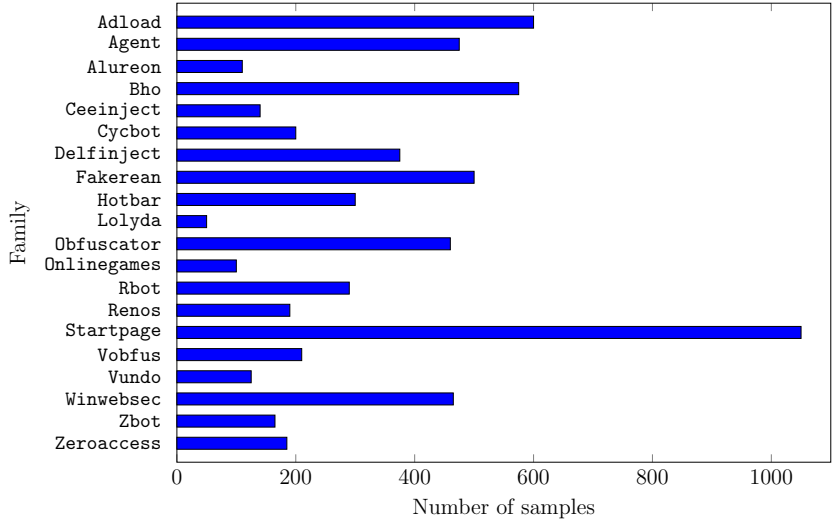
\begin{figure}[!htb]
\centering
\begin{tikzpicture}[scale=0.7, every node/.style={scale=1.0},rotate=-90]
    \begin{axis}[
        width  = 0.7*\textwidth,
        height = 14cm,
        ymin=0,ymax=1100,
        ytick={0,200,400,600,800,1000},
        major x tick style = transparent,
        ybar=5*\pgflinewidth,
        bar width=6.5pt,
        xlabel = {Family},
        xlabel style={rotate=180},
        ylabel = {Number of samples},
        yticklabel pos=right,
        symbolic x coords={
    Adload,
    Agent,
    Alureon, 
    Bho, 
    Ceeinject, 
    Cycbot, 
    Delfinject, 
    Fakerean, 
    Hotbar, 
    Lolyda, 
    Obfuscator, 
    Onlinegames, 
    Rbot, 
    Renos, 
    Startpage, 
    Vobfus, 
    Vundo, 
    Winwebsec, 
    Zbot, 
    Zeroaccess
	},
        xtick={
    Adload, 
    Agent, 
    Alureon, 
    Bho, 
    Ceeinject, 
    Cycbot, 
    Delfinject, 
    Fakerean, 
    Hotbar, 
    Lolyda, 
    Obfuscator, 
    Onlinegames, 
    Rbot, 
    Renos, 
    Startpage, 
    Vobfus, 
    Vundo, 
    Winwebsec, 
    Zbot, 
    Zeroaccess
	},
	y tick label style={
		rotate=90,
    		/pgf/number format/.cd,
		1000 sep={},
   		fixed,
   		fixed zerofill,
    		precision=0},
        x tick label style={
        		rotate=90,
		font=\small\tt,
		},
        enlarge x limits=0.045,
    ]
\addplot[fill=blue,opacity=1.00] 
coordinates {
(Adload,600) 
(Agent,475) 
(Alureon,110) 
(Bho,575) 
(Ceeinject,140) 
(Cycbot,200) 
(Delfinject,375) 
(Fakerean,500) 
(Hotbar,300) 
(Lolyda,50) 
(Obfuscator,460) 
(Onlinegames,100) 
(Rbot,290) 
(Renos,190) 
(Startpage,1050) 
(Vobfus,210) 
(Vundo,125) 
(Winwebsec,465) 
(Zbot,165) 
(Zeroaccess,185)
};
\end{axis}
\end{tikzpicture}
\caption{MalExe average size per family} 
\label{fig:average_size}
\end{figure}

The average file size per family is given in Figure~\ref{fig:average_size}. 
We observe that, on average, for example, Lolyda has an average 
file size of 35KB, while the average file size of Adload is 602KB and 
Startpage has the biggest average size at 1042KB. 


For our machine learning 
experiments, we need fixed-size input. 
There are multiple ways we can preprocess the data to achieve this. One 
reasonable approach is to extract a fixed amount of bytes and filter out 
smaller files, as is done in the papers~\cite{jain2019image,nagaraju2021auxiliaryclassifier}. 
Another approach is to have variable sizes, then 
resize to the desired width and height. 
In Section~\ref{sect:imageExtract}, we discuss image extraction from executable files 
and we experiment with different file sizes, as well as grayscale and color images. 

We divide this discussion of our implementation into four parts: dataset overview, 
image and feature extraction, data processing, and model hyperparameter tuning.
We employ Google Colab Pro+ to utilize Google's computing power to train multiple 
models on large datasets. Table~\ref{google_colab} shows the runtime and memory 
settings for each model. Note that for XGBoost, we utilize GPUs for faster training time. 

\begin{table}[!htb]
    \caption{Runtime environment specifications\label{google_colab}}
    \centering
    \adjustbox{scale=0.85}{
    \begin{tabular}{c|cc}\midrule\midrule
            Models & Runtime & Memory\\
    \midrule
            AC-GAN & TPU & 35GB \\
            DC-GAN & TPU & 35GB\\
            RBM & TPU & 35GB \\
            XGBoost & GPU & 51GB \\
            SVM & CPU & 51GB \\ 
            RF & CPU & 51GB \\ 
            KNN & CPU & 51GB \\ 
            MLP & CPU & 51GB \\ 
            Resnet152 & TPU/CPU & 35GB \\ 
    \midrule\midrule
    \end{tabular}
    }
\end{table}

\subsection{Image Extraction}\label{sect:imageExtract}

The sizes of the executable files vary and, as mentioned
above, we require fixed-size input for our learning experiments  
We consider two distinct approaches to generate fixed-size images. 
Our first approach is based on resizing,  
while for the second, we simply truncate.

For our resizing approach, we first divide the samples into 
bins and set a corresponding image width per bin. For each bin, 
a fixed width and variable height is used based on the sizes
given in~\cite{VASAN2020107138}. 
Table~\ref{width_table} shows the details of the bins and width.

\begin{table}[!htb]
\caption{Image width based on file size\label{width_table}}
\centering
\adjustbox{scale=0.85}{
\begin{tabular}{cc|cc}\midrule\midrule
File Size & Image Width & File Size & Image Width \\ \midrule
\zz0KB -- \zz10KB & \zz32 & 100KB -- \zz200KB & \zz384 \\
10KB -- \zz30KB & \zz64 & 200KB -- \zz500KB & \zz512 \\
30KB -- \zz60KB & 128 & 500KB -- 1000KB & \zz768 \\
60KB -- 100KB & 256 & $\hbox{}>\hbox{}$1000KB & 1024 \\ \midrule\midrule
\end{tabular}
}
\end{table}

After generating images from executable files
as described in the previous paragraph, we resize all images 
to~$128\times 128$. We refer to this image-generation approach
as our resizing method.
 
As an alternative approach for generating images from \texttt{exe} files,
we simply truncate the malware samples to the desired size,
with 0-paddings for smaller files. Thus, for~$128\times 128$ 
images, we only use the first~16,384 bytes of each executable files. 
We refer to this image generation approach as the truncating method. 

\subsubsection{Grayscale Images}

\begin{figure}[!htb]
    \centering
    \begin{tabular}{ccc}
        \includegraphics[width=0.32\textwidth]{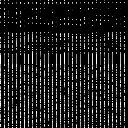}
& &
        \raisebox{0.5in}{\includegraphics[width=0.16\textwidth]{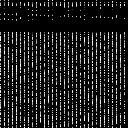}}
\\
        (a) Adload (256x256)
& &
        (b) Startpage (128x128) 
\\
    \end{tabular}
    \caption{Examples of grayscale malware images}\label{fig:grayscale}
\end{figure}

Grayscale images are straightforward to generate from executable files---we 
simply interpret the bytes as pixels in an image. This is the most popular approach in the
research literature for generating malware images; see,
for example~\cite{jain2019image,nagaraju2021auxiliaryclassifier, XIAO2021102420}. 
Only the byte sequence of the executable file is needed, 
and the processing is fast, even if resizing is employed. 
Examples of such grayscale malware images from our dataset
are given in Figure~\ref{fig:grayscale}.

\subsubsection{Color Images Using Color Map}

\begin{figure}[!htb]
\centering
\includegraphics[width=0.45\textwidth]{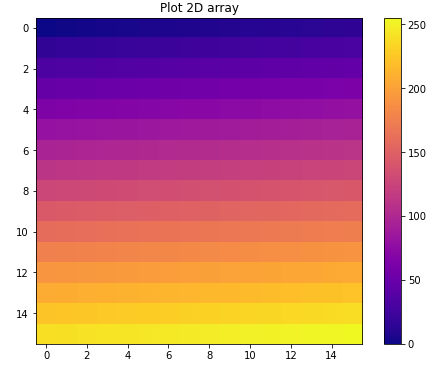}
\caption{Plasma colormap} 
\label{fig:plasma}
\end{figure}

In this section, we discuss generating color images from 
executable files using color map, as in~\cite{SINGH2019,VASAN2020107138}. First, 
we need to generate a 2D color map, which is a $16\times 16$ array where each 
element corresponds to an RGB value. There are a total~256 colors in this 
palette so we extract from the byte sequence a byte or 8-bit vector. We then 
split the byte into two parts, the first half of the byte represents the 
$y$-coordinate and the second half is the $x$-coordinate. Then we use these
coordinates to obtain the RGB value from the color map. Specifically, we use 
the ``plasma'' colormap, as given in
Figure~\ref{fig:plasma}. 

\begin{figure}[!htb]
\centering
\begin{tabular}{ccc}
    \includegraphics[width=0.1125\textwidth]{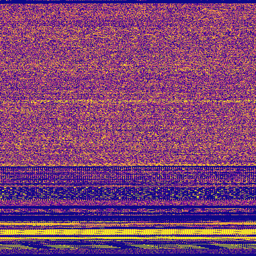}
    \includegraphics[width=0.1125\textwidth]{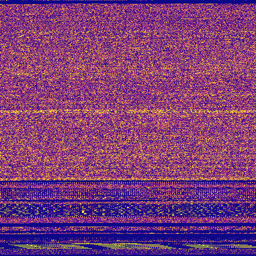}
    \includegraphics[width=0.1125\textwidth]{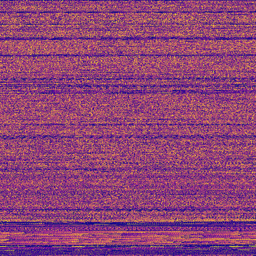}
& &
    \includegraphics[width=0.1125\textwidth]{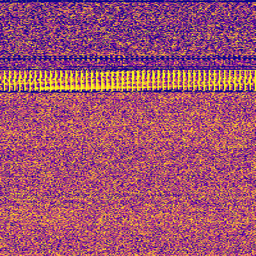}
    \includegraphics[width=0.1125\textwidth]{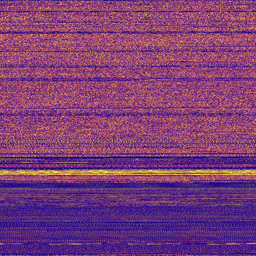}
    \includegraphics[width=0.1125\textwidth]{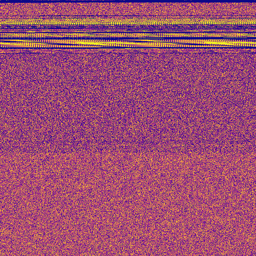}
\\
    (a) Adload
& &
    (b) Startpage
\\
\\[-1.5ex]
    \multicolumn{3}{c}{%
    \includegraphics[width=0.1125\textwidth]{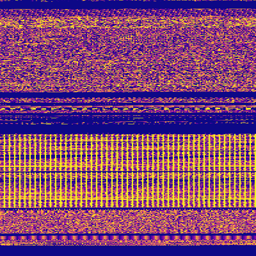}
    \includegraphics[width=0.1125\textwidth]{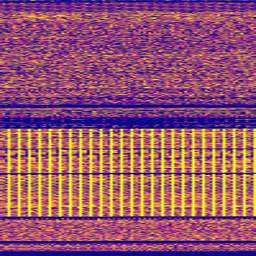}
    \includegraphics[width=0.1125\textwidth]{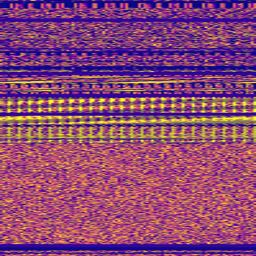}
    }
\\
    \multicolumn{3}{c}{(c) Ceeinject}
\end{tabular}
\vglue-0.05in
\caption{Colormap images of different families}\label{fig:images_colormap}
\end{figure}

After we generate the images using the color map, the images have 
different sizes based on the file sizes. We then resize all images to the 
fixed $128\times 128$ as in the examples in Figure~\ref{fig:images_colormap}. Each sample is now 
represented by an array of (128,128,3), and the data is ready to feed into 
machine learning models for training. We refer to this method as the colormap method.

\subsubsection{Color Images Using Three Consecutive Bytes}

\begin{figure}[!htb]
    \centering
    \begin{tabular}{ccc}
        \includegraphics[width=0.1125\textwidth]{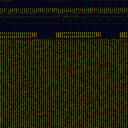}
        \includegraphics[width=0.1125\textwidth]{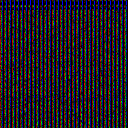}
        \includegraphics[width=0.1125\textwidth]{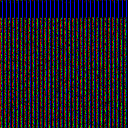}
& &
        \includegraphics[width=0.1125\textwidth]{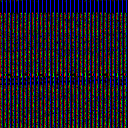}
        \includegraphics[width=0.1125\textwidth]{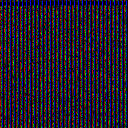}
        \includegraphics[width=0.1125\textwidth]{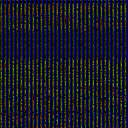}
\\
(a) Rbot
& & 
(b) Startpage
\\
\\[-1.5ex]
        \multicolumn{3}{c}{%
        \includegraphics[width=0.1125\textwidth]{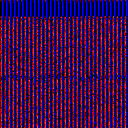}
        \includegraphics[width=0.1125\textwidth]{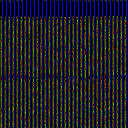}
        \includegraphics[width=0.1125\textwidth]{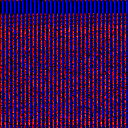}
        }
\\
        \multicolumn{3}{c}{(c) Ceeinject}
    \end{tabular}
    \vglue-0.05in
    \caption{Three-gram images of different families}\label{fig:images_3grams}
    \end{figure}

Another method for generating color images that we consider is to let
three consecutive byte values correspond to the~R, G and~B layers.
of an RGB image. We found that, visually,
the resulting images were not as colorful as expected. Therefore, we 
kept the red and green layers, but replaced the blue value with
$$
    \mbox{blue} = 255 - \mbox{ByteValue} .
$$
The resulting 3-grams images tend to have a blue background so 
that we can more easily distinguish 
them from grayscale images. Examples of this type of malware-derived color image 
are given in Figure~\ref{fig:images_3grams}.
We refer to this method of generating color images as  the ``3-grams'' approach.

Next, we discuss our third (and final) method for generating color images.
Then we conclude this section with a discussion of data pre-processing
and model hyperparameter tuning.

\subsubsection{Color Images from the PE File Format}

Fu et al.~\cite{section_colored} proposed to transform malware executables 
into color images based on the Windows PE file format. 
For each PE section, the entropy value is calculated once and used as the~R 
layer for the whole section. 
The~B layer is represented using the relative size of a section, while the~G layer
is just the byte values (as in grayscale images). We experimented with this method to extract colorful 
images from malware files. Entropy values and size ratios are calculated and scaled to 
the range of~0 to~255. The formulas for red and blue layers are
$$
\mbox{RedLayer} = \mbox{Entropy}\times \frac{255}{8}\mbox{\ \ and\ \ }
    \mbox{BlueLayer} = \frac{\mbox{SectionSize}}{\mbox{FileSize}}\times 255 .
$$

\begin{figure}[!htb]
    \centering
    \begin{tabular}{ccc}
        \includegraphics[width=0.1125\textwidth]{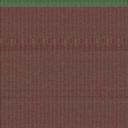}
        \includegraphics[width=0.1125\textwidth]{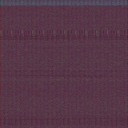}
        \includegraphics[width=0.1125\textwidth]{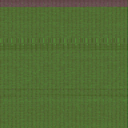}
& &
        \includegraphics[width=0.1125\textwidth]{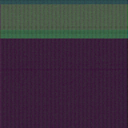}
        \includegraphics[width=0.1125\textwidth]{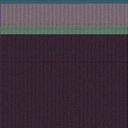}
        \includegraphics[width=0.1125\textwidth]{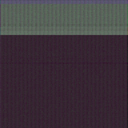}
\\
    (a) Adload
& &
    (b) Startpage
\\
\\[-1.5ex]
        \multicolumn{3}{c}{%
        \includegraphics[width=0.1125\textwidth]{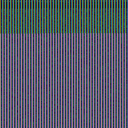}
        \includegraphics[width=0.1125\textwidth]{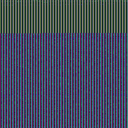}
        \includegraphics[width=0.1125\textwidth]{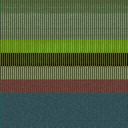}
        }
\\
    \multicolumn{3}{c}{(c) Ceeinject}
    \end{tabular}
    \vglue-0.05in
    \caption{PE images of different families}
    \label{fig:images_PE}
    \end{figure}

Figure~\ref{fig:images_PE} shows the PE images of samples from three 
families, namely, Adload, Startpage, and Ceeinject. Note that
Ceeinject uses extensive obfuscation, so we expect images from the Ceeinject 
family to differ more significantly than samples in the Adload or Startpage families

\subsection{Data Processing}

GANs require input data in the form of a 3-dimension array, while many other techniques
(e.g., SVM and XGBoost) expect 1-dimension array input. Therefore, we convert images
to vectors by flattening and scaling 
according to
$$
    \bar{x}= \frac{x-\mu}{\sigma}
$$
where $\mu$ is the mean and $\sigma$ is the standard deviation.
Data that is scaled to be between~0 and~1 may result in 
faster and more stable computation in XGBoost and RBM experiments.

\section{Experiments and Results}\label{chap:result}

In this section, we present and analyze our experimental results.
First, we consider model hyperparameter tuning, This is followed by
extensive multiclass experiments, where we compare
a wide variety of classification techniques, and we test each
of the four image generation methods introduced in 
Section~\ref{sect:imageExtract}, above. For our final set of experiments,
we analyze the generative capabilities of GAN-based techniques.

\subsection{Model Hyperparameter Tuning}

Figures~\ref{fig:acgan_generator} and~\ref{fig:acgan_discriminator} in
the Appendix show the AC-GAN generator and discriminator architectures,
respectively. Note that these architectures are specific to color
images of size~$128\times 128$, and that they are the same as 
those used in~\cite{acgan_implementation}. For other images, minor
modifications are required

For the AC-GAN generator, 
Manisha et al.~\cite{noise_dimension} showed that for color images with sizes of~$64\times 64$ 
and above, increasing the noise dimension has a significant positive impact on 
the generative images, and therefore both the discriminator and generator models would 
benefit. For this reason, we choose the noise dimension vectors for the generator to be
size~$(1000,1)$ instead of the more typical~$(100,1)$ 
for images of size $28\times 28$~\cite{noise_dimension}.

We perform hyperparameter tuning on each of our models using a grid search. 
The hyperparameters tested and selected for our RF, $k$-NN, MLP, SVM,
and XGBoost models are given in Tables~\ref{tab:parameters}~(a) through~(e),
respectively. For example, for
RF, we found that~600 estimators, the \texttt{entropy} function, and 
a maximum depth of~6 yielded the best results,
while for SVM, the \texttt{rbf} kernel with~$C=1$ was best, and so on.

\begin{table}[!htb]
\caption{Hyperparameters tested and selected}\label{tab:parameters}
    \centering
    \begin{tabular}{c}
(a) Random forest hyperparameters
\\
    \adjustbox{scale=0.775}{
    \begin{tabular}{c|ccc}\midrule\midrule
        Parameters & Description & Tested & Selected \\
    \midrule
        \texttt{nestimators} & Number of estimators & 100,200,400,600 & 600\\ 
        \texttt{criterion} & Function to measure quality & gini, entropy & entropy\\
        \texttt{maxdepth} & Maximum depth & 3,4,5,6 & 6\\
    \midrule\midrule
    \end{tabular}
    }
\\ \\
(b) $k$-NN hyperparameters
\\
    \adjustbox{scale=0.775}{
    \begin{tabular}{c|ccc}\midrule\midrule
        Parameters & Description & Tested & Selected \\
    \midrule
        \texttt{nneighbors} & Number of neighbors & 5,10,20,40 & 20\\ 
        \texttt{weights} & Function used in prediction & uniform, distance & distance\\
    \midrule\midrule
    \end{tabular}
    }
\\ \\
(c) MLP hyperparameters
\\
    \adjustbox{scale=0.775}{
    \begin{tabular}{c|ccc}\midrule\midrule
        Parameters & Description & Tested & Selected \\
    \midrule
        \texttt{hiddenlayer} & Sizes & (100,100,20), (100,100,100,20) & (100,100,100,20)\\ 
        \texttt{activation} & Activation function & logistic, tanh, relu & relu\\
        \texttt{alpha} & L2 penalty & 0.0001, 0.001, 0.01 & 0.0001\\
    \midrule\midrule
    \end{tabular}
    }
\\ \\
(d) SVM hyperparameters
\\
    \adjustbox{scale=0.775}{
    \begin{tabular}{c|ccc}\midrule\midrule
        Parameters & Description & Tested & Selected \\
    \midrule
        \texttt{kernel} & Kernel function & rbf, linear, poly & rbf\\ 
        $C$ & Regularization parameter & 1,10,100 & 1\\
    \midrule\midrule
    \end{tabular}
    }
\\ \\
(e) XGB hyperparameters
\\
    \adjustbox{scale=0.775}{
    \begin{tabular}{c|ccc}\midrule\midrule
        Parameters & Description & Tested & Selected \\
    \midrule
        \texttt{maxdepth} & Maximum depth & 4,5,6,7 & 6\\ 
        \texttt{learningrate} & Learning rate & 0.01, 0.02, 0.03 & 0.02\\
        \texttt{nestimators} & Number of estimators & 200,400,600 & 600\\
    \midrule\midrule
    \end{tabular}
    }
\end{tabular}
\end{table}

\subsection{Multiclass Classification}

We compare the performance of AC-GAN, RBM, SVM, and XGBoost on 
the four image extraction methods discussed above. Then we 
select the best of these image extraction method to compare GANs with 
other popular machine learning models. This latter comparison is with respect to 
a challenging multiclass malware classification problem.

\subsubsection{Image Extraction Comparison}\label{sect:image_ext}

\begin{figure}[!htb]
    \centering
    \begin{tikzpicture}[scale=0.8, every node/.style={scale=1.0}]
    \begin{axis}[
        width  = 0.825*\textwidth,
        height = 9.5cm,
        ymin=0.0,ymax=1.075,
        ytick={0,0.2,0.4,0.6,0.8,1.0},
        major x tick style = transparent,
        ybar=5*\pgflinewidth,
        bar width=11.0pt,
        xlabel = {Image conversion method},
        ylabel = {Accuracy},
        symbolic x coords={grayscale,colormap,PE,3-gram},
        xtick=data,
	y tick label style={
    		/pgf/number format/.cd,
   		fixed,
   		fixed zerofill,
    		precision=2},
        x tick label style={
		font=\small,
		},
        nodes near coords,
        every node near coord/.append style={rotate=90, 
        								   anchor=west,
								   font=\footnotesize,
								   /pgf/number format/.cd,
								   	fixed zerofill,
									precision=4
								   },
        enlarge x limits=0.175,
        legend cell align=left,
        legend pos=south east,
    ]
\addplot[fill=blue,opacity=1.00] 
coordinates {
(grayscale,0.7000)
(colormap,0.8400)
(PE,0.5742)
(3-gram,0.7248)
};
\addplot[fill=red,opacity=1.00] 
coordinates {
(grayscale,0.8000)
(colormap,0.8522)
(PE,0.6677)
(3-gram,0.8058)
};
\addplot[fill=brown,opacity=1.00] 
coordinates {
(grayscale,0.4518)
(colormap,0.4751)
(PE,0.4523)
(3-gram,0.4867)
};
\addplot[fill=green,opacity=1.00] 
coordinates {
(grayscale,0.8000)
(colormap,0.8944)
(PE,0.7100)
(3-gram,0.8488)
};
\legend{AC-GAN,SVM,RBM,XGBoost}
\end{axis}
\end{tikzpicture}
    \caption{Accuracies for malware families classification} 
    \label{fig:multi_accuracy_score}
\end{figure}
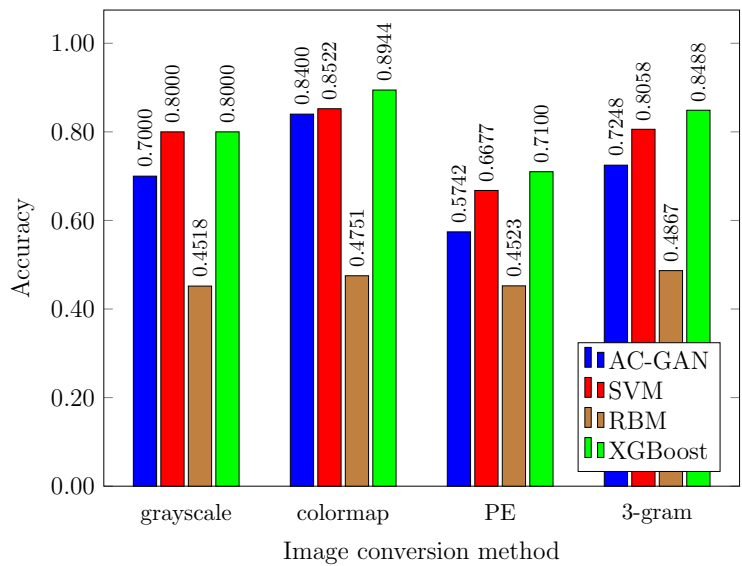

Figure~\ref{fig:multi_accuracy_score} provides our results for each of the four
image extraction techniques considered---grayscale, colormap, PE, and 3-gram---using 
each of the classifiers AC-GAN, RBM, SVM, and XGBoost. 
We observe that the PE features perform
the worst, while the color 3-grams sightly outperform grayscale images, with
the truncated colormap approach being the best overall. 
XGBoost, outperforms the other learning techniques in all cases,
but we will conduct a much more thorough comparison of learning
techniques in the next section.


In summary, we have determined that the truncated colormap method provides
the best performance among the image generation techniques tested.
In the next section,
we compare a wide variety of learning techniques based on this truncated colormap method.
For the colormap experiments in the next section, 
we only consider the first~$128\times 128$ 
bytes of the executables;
if the file file size is smaller, we pad with~$0$ to the necessary size.

\subsubsection{Machine Learning Model Comparison}

\begin{figure}[!htb]
    \centering
        \includegraphics[width=0.4\textwidth]{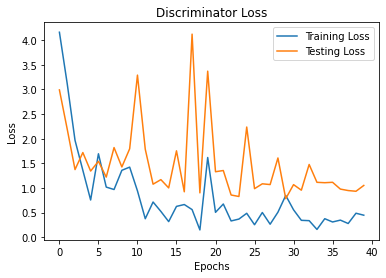}
        \includegraphics[width=0.4\textwidth]{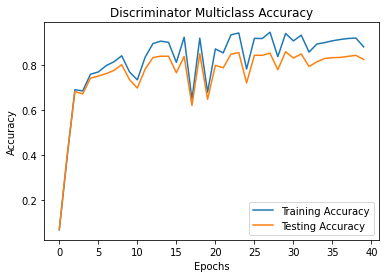}
    \caption{AC-GAN discriminator loss and accuracy}
    \label{fig:acgan_loss_accuracy}
\end{figure}

Figure~\ref{fig:acgan_loss_accuracy} shows the accuracy and loss graphs 
for both training and testing of the AC-GAN discriminator. After~30 epochs, there is not 
much improvement for the discriminator, and therefore in all subsequent
experiments, we train the AC-GAN discriminator 
for~30 epochs. 

Malware classification results for all of the classifiers we tested
are given in Table~\ref{best_results}.
From this table, we see that Resnet152 performs the best, with XGBoost, MLP,
and AC-GAN all yielding reasonably strong results.

\begin{table}[!htb]
    \caption{Results of various models on truncated colormap images}\label{best_results}
    \centering
    \adjustbox{scale=0.8}{
    \begin{tabular}{c|ccccc}\midrule\midrule
        Models & Type & Accuracy & Precision & Recall & F1-Score \\ \midrule
        $k$-NN & Maching Learning & 76.94\% & 87\% & 77\% & 79\% \\
        SVM & Machine Learning &  85.22\% & 88\% & 86\% & 86\% \\ 
        MLP & Machine Learning & 86.97\% & 86\% & 86\% & 86\% \\
        RF & Ensemble & 72.56\% & 80\% & 69\% & 70\% \\
        XGBoost & Ensemble & 89.44\% & 90\% & 89\% & 89\% \\ 
        AC-GAN & Deep Learning & 84.00\% & 86\% & 86\% & 85\% \\
        Resnet152 & Deep Residual Network& 91.39\% & 91\% & 91\% & 91\% \\
    \midrule\midrule
    \end{tabular}
    }
\end{table}

\begin{figure}[!htb]
    \centering
\begin{tikzpicture}[scale=0.75]
\begin{axis}[ 
		   width=0.8\textwidth,
		   height=0.575\textwidth,
	 	   x tick label style={
   		 	/pgf/number format/.cd,
			/pgf/number format/1000 sep={},
   			fixed,
   			fixed zerofill,
    			precision=0
		   },
	 	   y tick label style={
    		 	/pgf/number format/.cd,
   			fixed,
   			fixed zerofill,
    			precision=2
		    },
                    xmin=0,xmax=620,
                    ymin=0.70,ymax=1.00,
                    legend pos=south west,
                    xtick={0,100,200,300,400,500,600},
                    ytick={0.70,0.80,0.90,1.00},
                    ylabel={Accuracy},
                    xlabel={Minutes}] 
\addplot[only marks,color=blue,ultra thick,mark=*,mark size=2.0] coordinates { 
(10,0.7250)
(10,0.7700)
(20,0.8750)
(50,0.9200)
(85,0.8900)
(255,0.8600)
(590,0.8400)
};
\node[anchor=west] at (axis cs:20,0.7250) {RF};
\node[anchor=west] at (axis cs:20,0.7700) {$k$-NN};
\node[anchor=west] at (axis cs:0,0.8550) {MLP};
\node[anchor=west] at (axis cs:5,0.9400) {ResNet152};
\node[anchor=west] at (axis cs:95,0.8900) {XGBoost};
\node[anchor=west] at (axis cs:265,0.8600) {SVM};
\node[anchor=east] at (axis cs:580,0.8400) {AC-GAN};
\end{axis}
\end{tikzpicture}
    \caption{Accuracy vs training time comparison}
    \label{fig:accuracy_vs_time}
\end{figure}
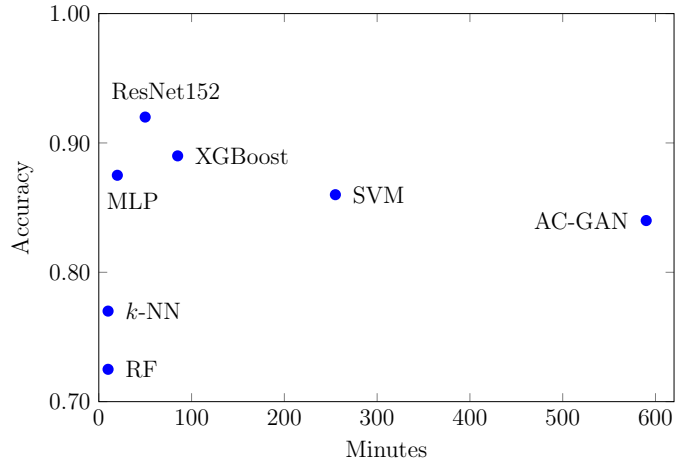

Figure~\ref{fig:accuracy_vs_time} compares the training time and performance of the various models. 
MLP and XGBoost perform exceptional well considering the training time is less 
than some of the other deep learning models, such as AC-GAN. Resnet152 is 
clearly our best model, as it gives the best performance, and the training time
is fast due to it being pre-trained. AC-GAN takes approximately~500 seconds per epoch, or 
around 4 hours to finish training; SVM takes 3 hours for training and 2 hours 
for prediction. All of our other models are relatively fast, with each 
taking less than an hour to train. 


\begin{figure}[!htb]
    \centering
\begin{tikzpicture}[scale=0.5]
    \begin{axis}[
        width=20cm,
        height=20cm,
	colormap={bluewhite}{color=(white) rgb255=(100,149,237)},
        xticklabels={
    Adload,
    Agent,
    Alureon, 
    Bho, 
    Ceeinject, 
    Cycbot, 
    Delfinject, 
    Fakerean, 
    Hotbar, 
    Lolyda, 
    Obfuscator, 
    Onlinegames, 
    Rbot, 
    Renos, 
    Startpage, 
    Vobfus, 
    Vundo, 
    Winwebsec, 
    Zbot, 
    Zeroaccess
        },
        xtick={0,...,19},
        xtick style={draw=none},
	xticklabel style={anchor=east,rotate=60,yshift=-5pt,font=\tt},
        yticklabels={
    Adload,
    Agent,
    Alureon, 
    Bho, 
    Ceeinject, 
    Cycbot, 
    Delfinject, 
    Fakerean, 
    Hotbar, 
    Lolyda, 
    Obfuscator, 
    Onlinegames, 
    Rbot, 
    Renos, 
    Startpage, 
    Vobfus, 
    Vundo, 
    Winwebsec, 
    Zbot, 
    Zeroaccess
        },
        ytick={0,...,19},
        ytick style={draw=none},
        enlargelimits=false,
        yticklabel style={font=\tt},
        colorbar,
        colorbar style={
            ytick={0.0,0.2,0.4,0.6,0.8,1.0},
            yticklabels={0.0,0.2,0.4,0.6,0.8,1.0},
            yticklabel={\pgfmathprintnumber\tick},
            yticklabel style={
            		/pgf/number format/fixed,
			/pgf/number format/precision=1}
        },
        point meta min=0.0,
        point meta max=1.0,
        nodes near coords={\pgfmathprintnumber\pgfplotspointmeta},
        nodes near coords black white/.style={
            small value/.style={
                yshift=-7pt,
                text=black,
                /pgf/number format/fixed,
                /pgf/number format/precision=2,
                /pgf/number format/zerofill=true,
            },
            large value/.style={
                yshift=-7pt,
                text=white,
                /pgf/number format/fixed,
                /pgf/number format/precision=2
            },
            every node near coord/.style={
                check for zero/.code={
                    \pgfmathfloatifflags{\pgfplotspointmeta}{0}{
                        \pgfkeys{/tikz/coordinate}
                    }{
                        \begingroup
                        \pgfkeys{/pgf/fpu}
                        \pgfmathparse{\pgfplotspointmeta<#1}
                        \global\let\result=\pgfmathresult
                        \endgroup
                        %
                        %
                        \pgfmathfloatcreate{1}{1.0}{0}
                        \let\ONE=\pgfmathresult
                        \ifx\result\ONE
                            \pgfkeysalso{/pgfplots/small value}
                        \else
                            \pgfkeysalso{/pgfplots/large value}
                        \fi
                    }
                },
                check for zero,
            },
        },
        nodes near coords black white=0.5,
    ]
        \addplot[
            matrix plot,
            mesh/cols=20,
            point meta=explicit,draw=gray
        ] table [meta=C] {
            x y C
0	0	0.97
1	0	0
2	0	0
3	0	0
4	0	0
5	0	0
6	0	0
7	0	0
8	0	0.01
9	0	0
10	0	0
11	0	0
12	0	0
13	0	0
14	0	0
15	0	0
16	0	0
17	0	0
18	0	0
19	0	0
0	1	0
1	1	0.77
2	1	0.01
3	1	0
4	1	0.01
5	1	0
6	1	0.03
7	1	0
8	1	0.01
9	1	0
10	1	0.04
11	1	0.04
12	1	0.03
13	1	0.01
14	1	0.03
15	1	0
16	1	0.02
17	1	0.01
18	1	0.01
19	1	0
0	2	0
1	2	0
2	2	0.84
3	2	0
4	2	0.01
5	2	0
6	2	0
7	2	0
8	2	0
9	2	0
10	2	0.02
11	2	0
12	2	0.02
13	2	0.01
14	2	0
15	2	0
16	2	0.05
17	2	0
18	2	0.02
19	2	0
0	3	0
1	3	0
2	3	0
3	3	0.97
4	3	0
5	3	0
6	3	0
7	3	0
8	3	0
9	3	0
10	3	0
11	3	0
12	3	0
13	3	0
14	3	0
15	3	0
16	3	0.01
17	3	0
18	3	0
19	3	0
0	4	0
1	4	0
2	4	0.01
3	4	0
4	4	0.87
5	4	0.01
6	4	0.01
7	4	0.01
8	4	0
9	4	0
10	4	0.02
11	4	0.01
12	4	0.03
13	4	0.01
14	4	0
15	4	0
16	4	0.02
17	4	0.01
18	4	0.01
19	4	0
0	5	0
1	5	0
2	5	0
3	5	0
4	5	0
5	5	0.93
6	5	0
7	5	0
8	5	0
9	5	0.01
10	5	0
11	5	0.01
12	5	0.02
13	5	0
14	5	0
15	5	0
16	5	0.02
17	5	0
18	5	0
19	5	0
0	6	0
1	6	0.01
2	6	0
3	6	0
4	6	0
5	6	0
6	6	0.88
7	6	0.01
8	6	0
9	6	0
10	6	0.04
11	6	0
12	6	0.02
13	6	0
14	6	0
15	6	0
16	6	0.01
17	6	0
18	6	0.01
19	6	0
0	7	0
1	7	0
2	7	0.01
3	7	0
4	7	0.01
5	7	0.01
6	7	0.01
7	7	0.86
8	7	0.01
9	7	0
10	7	0.04
11	7	0
12	7	0.03
13	7	0
14	7	0
15	7	0
16	7	0.03
17	7	0
18	7	0.01
19	7	0
0	8	0
1	8	0
2	8	0
3	8	0
4	8	0
5	8	0
6	8	0
7	8	0
8	8	1
9	8	0
10	8	0
11	8	0
12	8	0
13	8	0
14	8	0
15	8	0
16	8	0
17	8	0
18	8	0
19	8	0
0	9	0
1	9	0
2	9	0
3	9	0
4	9	0
5	9	0
6	9	0
7	9	0
8	9	0
9	9	1
10	9	0
11	9	0
12	9	0
13	9	0
14	9	0
15	9	0
16	9	0
17	9	0
18	9	0
19	9	0
0	10	0
1	10	0.03
2	10	0.03
3	10	0
4	10	0.01
5	10	0.02
6	10	0.05
7	10	0.01
8	10	0.01
9	10	0
10	10	0.56
11	10	0
12	10	0.11
13	10	0.03
14	10	0
15	10	0
16	10	0.08
17	10	0.01
18	10	0.04
19	10	0
0	11	0
1	11	0.03
2	11	0
3	11	0
4	11	0.01
5	11	0
6	11	0.01
7	11	0.01
8	11	0
9	11	0
10	11	0.01
11	11	0.88
12	11	0.02
13	11	0
14	11	0
15	11	0
16	11	0.02
17	11	0
18	11	0
19	11	0
0	12	0
1	12	0
2	12	0
3	12	0
4	12	0.01
5	12	0
6	12	0.03
7	12	0.01
8	12	0
9	12	0
10	12	0.06
11	12	0
12	12	0.79
13	12	0.01
14	12	0
15	12	0
16	12	0.03
17	12	0
18	12	0.03
19	12	0
0	13	0
1	13	0
2	13	0
3	13	0.01
4	13	0.01
5	13	0.01
6	13	0
7	13	0
8	13	0
9	13	0
10	13	0.01
11	13	0.01
12	13	0.01
13	13	0.91
14	13	0
15	13	0
16	13	0.02
17	13	0
18	13	0
19	13	0
0	14	0
1	14	0.11
2	14	0
3	14	0
4	14	0
5	14	0
6	14	0
7	14	0
8	14	0
9	14	0
10	14	0
11	14	0
12	14	0.01
13	14	0
14	14	0.84
15	14	0
16	14	0.01
17	14	0
18	14	0
19	14	0
0	15	0
1	15	0
2	15	0
3	15	0
4	15	0
5	15	0
6	15	0
7	15	0
8	15	0
9	15	0
10	15	0.01
11	15	0
12	15	0
13	15	0
14	15	0
15	15	0.98
16	15	0.01
17	15	0
18	15	0
19	15	0
0	16	0
1	16	0
2	16	0.01
3	16	0
4	16	0.01
5	16	0
6	16	0.01
7	16	0.01
8	16	0
9	16	0
10	16	0.01
11	16	0.01
12	16	0.02
13	16	0.01
14	16	0
15	16	0.01
16	16	0.89
17	16	0.01
18	16	0.01
19	16	0
0	17	0
1	17	0
2	17	0
3	17	0
4	17	0
5	17	0
6	17	0
7	17	0
8	17	0
9	17	0
10	17	0
11	17	0
12	17	0
13	17	0
14	17	0
15	17	0
16	17	0
17	17	1
18	17	0
19	17	0
0	18	0
1	18	0
2	18	0
3	18	0
4	18	0
5	18	0
6	18	0
7	18	0
8	18	0
9	18	0
10	18	0
11	18	0
12	18	0
13	18	0
14	18	0
15	18	0
16	18	0.01
17	18	0
18	18	0.99
19	18	0
0	19	0
1	19	0
2	19	0
3	19	0
4	19	0
5	19	0
6	19	0
7	19	0
8	19	0
9	19	0
10	19	0.01
11	19	0.01
12	19	0.01
13	19	0
14	19	0
15	19	0
16	19	0
17	19	0
18	19	0.01
19	19	0.96
         };
    \end{axis}
\end{tikzpicture}
%
    \caption{Confusion matrix for Resnet152 using colormap method} 
    \label{fig:resnet_truncated_cm}
\end{figure}
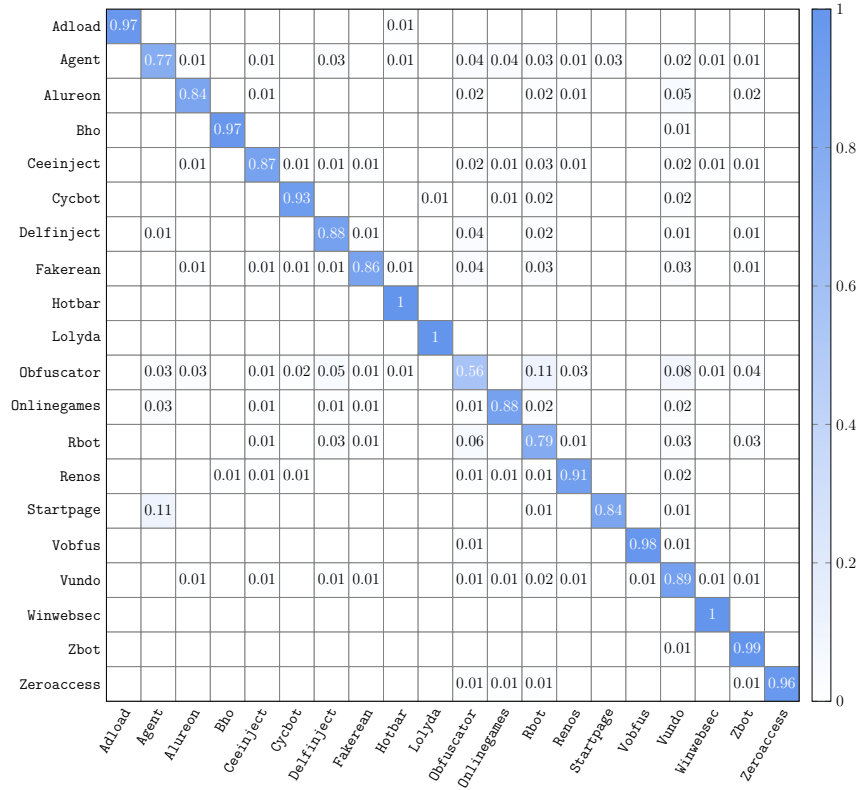

As ResNet152 is our best performer, we provide its confusion matrix 
in Figure~\ref{fig:resnet_truncated_cm}. We can see that the three families 
that are causing the most classification problems for this model are Obfuscator, Rbot, and 
Agent. Since Obfuscator 
obfuscates its code, we expect this family
to be a difficult classification problem. Agent is a general family that includes 
malware with multiple purposes, and it also makes sense that machine learning models 
would have difficulty classifying more general types. 


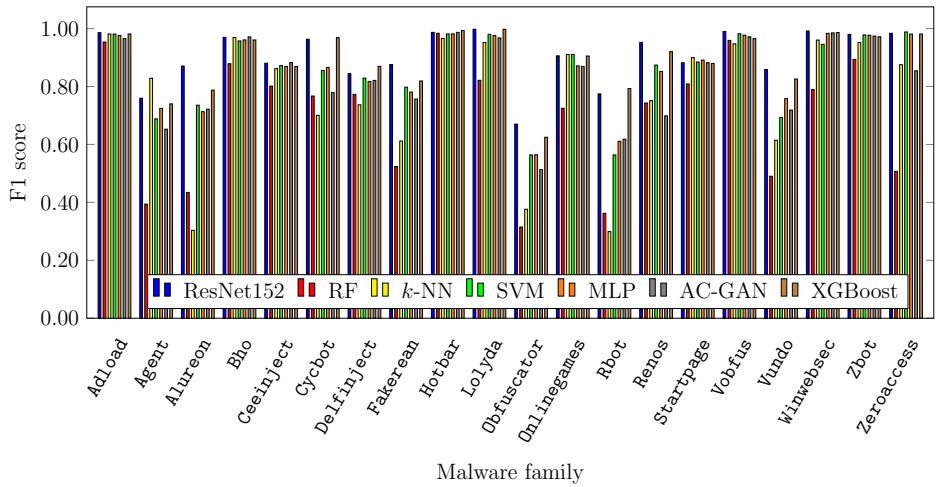
\begin{figure}[!htb]
    \centering
    \begin{tikzpicture}[scale=0.7, every node/.style={scale=1.0}]
    \begin{axis}[
        width  = 1.2*\textwidth,
        height = 7.5cm,
        ymin=0.0,ymax=1.075,
        ytick={0,0.2,0.4,0.6,0.8,1.0},
        major x tick style = transparent,
        ybar=3*\pgflinewidth,
        bar width=1.5pt,
        xlabel = {Malware family},
        ylabel = {F1 score},
        symbolic x coords={
    Adload,
    Agent,
    Alureon, 
    Bho, 
    Ceeinject, 
    Cycbot, 
    Delfinject, 
    Fakerean, 
    Hotbar, 
    Lolyda, 
    Obfuscator, 
    Onlinegames, 
    Rbot, 
    Renos, 
    Startpage, 
    Vobfus, 
    Vundo, 
    Winwebsec, 
    Zbot, 
    Zeroaccess
	},
        xtick={
    Adload, 
    Agent, 
    Alureon, 
    Bho, 
    Ceeinject, 
    Cycbot, 
    Delfinject, 
    Fakerean, 
    Hotbar, 
    Lolyda, 
    Obfuscator, 
    Onlinegames, 
    Rbot, 
    Renos, 
    Startpage, 
    Vobfus, 
    Vundo, 
    Winwebsec, 
    Zbot, 
    Zeroaccess
	},
	y tick label style={
    		/pgf/number format/.cd,
   		fixed,
   		fixed zerofill,
    		precision=2},
        x tick label style={
        		rotate=60,
		font=\small\tt,
		anchor=north east,
		},
        enlarge x limits=0.035,
        legend cell align=left,
        legend pos=south east,
        legend style={legend columns=-1,column sep=0.175cm},
    ]
\addplot[fill=blue,opacity=1.00] 
coordinates {
(Adload,0.985781991)
(Agent,0.759643917)
(Alureon,0.870689655)
(Bho,0.969432314)
(Ceeinject,0.880222841)
(Cycbot,0.963144963)
(Delfinject,0.84434968)
(Fakerean,0.875647668)
(Hotbar,0.986842105)
(Lolyda,0.997578692)
(Obfuscator,0.670068027)
(Onlinegames,0.905861456)
(Rbot,0.774509804)
(Renos,0.952380952)
(Startpage,0.881773399)
(Vobfus,0.989795918)
(Vundo,0.858681023)
(Winwebsec,0.991836735)
(Zbot,0.979811575)
(Zeroaccess,0.983682984)
};
\addplot[fill=red,opacity=1.00] 
coordinates {
(Adload,0.953545232)
(Agent,0.393939394)
(Alureon,0.434367542)
(Bho,0.878640777)
(Ceeinject,0.801324503)
(Cycbot,0.767195767)
(Delfinject,0.772983114)
(Fakerean,0.52398524)
(Hotbar,0.983766234)
(Lolyda,0.821428571)
(Obfuscator,0.31496063)
(Onlinegames,0.724960254)
(Rbot,0.362264151)
(Renos,0.742980562)
(Startpage,0.808716707)
(Vobfus,0.959183673)
(Vundo,0.490230906)
(Winwebsec,0.789242591)
(Zbot,0.893401015)
(Zeroaccess,0.507211538)
};
\addplot[fill=yellow,opacity=1.00] 
coordinates {
(Adload,0.981132075)
(Agent,0.828478964)
(Alureon,0.303514377)
(Bho,0.969162996)
(Ceeinject,0.86163522)
(Cycbot,0.700610998)
(Delfinject,0.736842105)
(Fakerean,0.611842105)
(Hotbar,0.966101695)
(Lolyda,0.951898734)
(Obfuscator,0.376190476)
(Onlinegames,0.910447761)
(Rbot,0.299003322)
(Renos,0.751111111)
(Startpage,0.899742931)
(Vobfus,0.947368421)
(Vundo,0.61452514)
(Winwebsec,0.96031746)
(Zbot,0.95212766)
(Zeroaccess,0.875331565)
};
\addplot[fill=green,opacity=1.00] 
coordinates {
(Adload,0.980952381)
(Agent,0.688259109)
(Alureon,0.735483871)
(Bho,0.957303371)
(Ceeinject,0.872274143)
(Cycbot,0.855614973)
(Delfinject,0.829059829)
(Fakerean,0.797687861)
(Hotbar,0.98163606)
(Lolyda,0.980295567)
(Obfuscator,0.563636364)
(Onlinegames,0.910746812)
(Rbot,0.563706564)
(Renos,0.87398374)
(Startpage,0.884353741)
(Vobfus,0.982097187)
(Vundo,0.692931634)
(Winwebsec,0.945383615)
(Zbot,0.978082192)
(Zeroaccess,0.988290398)
};
\addplot[fill=orange,opacity=1.00] 
coordinates {
(Adload,0.976303318)
(Agent,0.724458204)
(Alureon,0.712945591)
(Bho,0.960869565)
(Ceeinject,0.869300912)
(Cycbot,0.865882353)
(Delfinject,0.817460317)
(Fakerean,0.780722892)
(Hotbar,0.981697171)
(Lolyda,0.975961538)
(Obfuscator,0.564315353)
(Onlinegames,0.871186441)
(Rbot,0.610722611)
(Renos,0.852207294)
(Startpage,0.891139241)
(Vobfus,0.97721519)
(Vundo,0.757952974)
(Winwebsec,0.983628922)
(Zbot,0.97699594)
(Zeroaccess,0.981042654)
};
\addplot[fill=gray,opacity=1.00] 
coordinates {
(Adload,0.965034965)
(Agent,0.652631579)
(Alureon,0.721698113)
(Bho,0.971302428)
(Ceeinject,0.882716049)
(Cycbot,0.778656126)
(Delfinject,0.821052632)
(Fakerean,0.75659824)
(Hotbar,0.986885246)
(Lolyda,0.967741935)
(Obfuscator,0.513011152)
(Onlinegames,0.869230769)
(Rbot,0.618181818)
(Renos,0.698591549)
(Startpage,0.882205514)
(Vobfus,0.971722365)
(Vundo,0.718801997)
(Winwebsec,0.985074627)
(Zbot,0.97414966)
(Zeroaccess,0.854251012)
};
\addplot[fill=brown,opacity=1.00] 
coordinates {
(Adload,0.981308411)
(Agent,0.74)
(Alureon,0.7875)
(Bho,0.96069869)
(Ceeinject,0.869047619)
(Cycbot,0.968674699)
(Delfinject,0.869565217)
(Fakerean,0.819148936)
(Hotbar,0.993442623)
(Lolyda,0.997590361)
(Obfuscator,0.624787776)
(Onlinegames,0.905263158)
(Rbot,0.792838875)
(Renos,0.920454545)
(Startpage,0.879795396)
(Vobfus,0.965174129)
(Vundo,0.825806452)
(Winwebsec,0.986431479)
(Zbot,0.971659919)
(Zeroaccess,0.981308411)
};
\legend{ResNet152,RF,$k$-NN,SVM,MLP,AC-GAN,XGBoost}
\end{axis}
\end{tikzpicture}
    \caption{F1 scores for each family}\label{fig:f1_score}
\end{figure}

Figure~\ref{fig:f1_score} confirms that more general and heavily obfuscated 
families are hard to classify. These families include Agent, Ceeinject, Obfuscator, and Rbot,
which are consistently classified poorly by all learning techniques tested.
Resnet152 performs best on most of the families. Interestingly, 
although $k$-NN does not perform well overall, it performs better than any of
the other models on the challenging Agent family.


\subsubsection{Ensemble Classifiers}

In an attempt to improve the accuracy for the most difficult families, such as Agent or Rbot, 
we generate an ensemble based on all~7 of the classifiers discussed above: 
AC-GAN, $k$-NN, MLP, Resnet152, RF, SVM, and XGBoost. First, we consider a simple
voting procedure, in which case the ensemble gives us an accuracy of~91.60\%,
which is only a marginal improvement over the~91.39\%\ accuracy achieved by ResNet152 alone. 

\begin{figure}[!htb]
    \centering
\begin{tikzpicture}[scale=0.5]
    \begin{axis}[
        width=20cm,
        height=20cm,
	colormap={bluewhite}{color=(white) rgb255=(100,149,237)},
        xticklabels={
    Adload,
    Agent,
    Alureon, 
    Bho, 
    Ceeinject, 
    Cycbot, 
    Delfinject, 
    Fakerean, 
    Hotbar, 
    Lolyda, 
    Obfuscator, 
    Onlinegames, 
    Rbot, 
    Renos, 
    Startpage, 
    Vobfus, 
    Vundo, 
    Winwebsec, 
    Zbot, 
    Zeroaccess
        },
        xtick={0,...,19},
        xtick style={draw=none},
	xticklabel style={anchor=east,rotate=60,yshift=-5pt,font=\tt},
        yticklabels={
    Adload,
    Agent,
    Alureon, 
    Bho, 
    Ceeinject, 
    Cycbot, 
    Delfinject, 
    Fakerean, 
    Hotbar, 
    Lolyda, 
    Obfuscator, 
    Onlinegames, 
    Rbot, 
    Renos, 
    Startpage, 
    Vobfus, 
    Vundo, 
    Winwebsec, 
    Zbot, 
    Zeroaccess
        },
        ytick={0,...,19},
        ytick style={draw=none},
        enlargelimits=false,
        yticklabel style={font=\tt},
        colorbar,
        colorbar style={
            ytick={0.0,0.2,0.4,0.6,0.8,1.0},
            yticklabels={0.0,0.2,0.4,0.6,0.8,1.0},
            yticklabel={\pgfmathprintnumber\tick},
            yticklabel style={
            		/pgf/number format/fixed,
			/pgf/number format/precision=1}
        },
        point meta min=0.0,
        point meta max=1.0,
        nodes near coords={\pgfmathprintnumber\pgfplotspointmeta},
        nodes near coords black white/.style={
            small value/.style={
                yshift=-7pt,
                text=black,
                /pgf/number format/fixed,
                /pgf/number format/precision=2,
                /pgf/number format/zerofill=true,
            },
            large value/.style={
                yshift=-7pt,
                text=white,
                /pgf/number format/fixed,
                /pgf/number format/precision=2,
                /pgf/number format/zerofill=true,
            },
            every node near coord/.style={
                check for zero/.code={
                    \pgfmathfloatifflags{\pgfplotspointmeta}{0}{
                        \pgfkeys{/tikz/coordinate}
                    }{
                        \begingroup
                        \pgfkeys{/pgf/fpu}
                        \pgfmathparse{\pgfplotspointmeta<#1}
                        \global\let\result=\pgfmathresult
                        \endgroup
                        %
                        %
                        \pgfmathfloatcreate{1}{1.0}{0}
                        \let\ONE=\pgfmathresult
                        \ifx\result\ONE
                            \pgfkeysalso{/pgfplots/small value}
                        \else
                            \pgfkeysalso{/pgfplots/large value}
                        \fi
                    }
                },
                check for zero,
            },
        },
        nodes near coords black white=0.5,
    ]
        \addplot[
            matrix plot,
            mesh/cols=20,
            point meta=explicit,draw=gray
        ] table [meta=C] {
            x y C
 0 0 0.96
1 0 0.01
2 0 0
3 0 0
4 0 0
5 0 0
6 0 0
7 0 0
8 0 0
9 0 0
10 0 0.01
11 0 0
12 0 0
13 0 0
14 0 0
15 0 0
16 0 0
17 0 0
18 0 0
19 0 0
0 1 0
1 1 0.84
2 1 0.03
3 1 0
4 1 0
5 1 0
6 1 0.03
7 1 0.01
8 1 0
9 1 0
10 1 0.04
11 1 0.01
12 1 0.01
13 1 0.01
14 1 0.01
15 1 0
16 1 0.03
17 1 0
18 1 0
19 1 0
0 2 0
1 2 0.01
2 2 0.89
3 2 0
4 2 0
5 2 0
6 2 0
7 2 0
8 2 0
9 2 0
10 2 0.02
11 2 0
12 2 0
13 2 0
14 2 0
15 2 0
16 2 0.06
17 2 0
18 2 0
19 2 0
0 3 0
1 3 0
2 3 0
3 3 0.97
4 3 0
5 3 0
6 3 0
7 3 0
8 3 0
9 3 0
10 3 0.02
11 3 0
12 3 0
13 3 0
14 3 0
15 3 0
16 3 0
17 3 0
18 3 0
19 3 0
0 4 0
1 4 0.01
2 4 0.02
3 4 0
4 4 0.85
5 4 0.01
6 4 0.01
7 4 0.01
8 4 0
9 4 0
10 4 0.03
11 4 0.01
12 4 0.01
13 4 0
14 4 0
15 4 0
16 4 0.04
17 4 0.01
18 4 0
19 4 0
0 5 0
1 5 0
2 5 0.01
3 5 0
4 5 0
5 5 0.91
6 5 0
7 5 0
8 5 0
9 5 0
10 5 0.02
11 5 0
12 5 0
13 5 0
14 5 0
15 5 0
16 5 0.04
17 5 0
18 5 0
19 5 0
0 6 0
1 6 0
2 6 0.01
3 6 0
4 6 0
5 6 0
6 6 0.92
7 6 0
8 6 0
9 6 0
10 6 0.06
11 6 0
12 6 0
13 6 0
14 6 0
15 6 0
16 6 0
17 6 0
18 6 0
19 6 0
0 7 0
1 7 0
2 7 0.02
3 7 0
4 7 0
5 7 0.01
6 7 0
7 7 0.9
8 7 0
9 7 0
10 7 0.05
11 7 0
12 7 0.02
13 7 0
14 7 0
15 7 0
16 7 0.02
17 7 0
18 7 0
19 7 0
0 8 0
1 8 0
2 8 0
3 8 0
4 8 0
5 8 0
6 8 0
7 8 0
8 8 0.99
9 8 0
10 8 0
11 8 0
12 8 0
13 8 0
14 8 0
15 8 0
16 8 0.01
17 8 0
18 8 0
19 8 0
0 9 0
1 9 0
2 9 0
3 9 0
4 9 0
5 9 0
6 9 0
7 9 0
8 9 0
9 9 1
10 9 0
11 9 0
12 9 0
13 9 0
14 9 0
15 9 0
16 9 0
17 9 0
18 9 0
19 9 0
0 10 0
1 10 0.02
2 10 0.04
3 10 0
4 10 0
5 10 0.01
6 10 0.07
7 10 0.03
8 10 0
9 10 0
10 10 0.72
11 10 0
12 10 0.04
13 10 0.01
14 10 0
15 10 0
16 10 0.05
17 10 0.01
18 10 0
19 10 0
0 11 0
1 11 0.01
2 11 0.01
3 11 0
4 11 0
5 11 0
6 11 0.01
7 11 0.01
8 11 0
9 11 0
10 11 0.04
11 11 0.91
12 11 0.01
13 11 0
14 11 0
15 11 0
16 11 0
17 11 0
18 11 0
19 11 0
0 12 0
1 12 0
2 12 0
3 12 0
4 12 0
5 12 0
6 12 0.04
7 12 0.03
8 12 0
9 12 0
10 12 0.09
11 12 0
12 12 0.79
13 12 0
14 12 0
15 12 0
16 12 0.03
17 12 0
18 12 0
19 12 0
0 13 0
1 13 0
2 13 0.01
3 13 0
4 13 0
5 13 0.01
6 13 0
7 13 0
8 13 0
9 13 0
10 13 0.02
11 13 0
12 13 0
13 13 0.92
14 13 0
15 13 0
16 13 0.03
17 13 0
18 13 0
19 13 0
0 14 0
1 14 0.1
2 14 0
3 14 0
4 14 0
5 14 0
6 14 0.01
7 14 0.02
8 14 0
9 14 0
10 14 0.02
11 14 0
12 14 0.01
13 14 0
14 14 0.82
15 14 0
16 14 0.02
17 14 0
18 14 0
19 14 0
0 15 0
1 15 0
2 15 0.01
3 15 0
4 15 0
5 15 0.01
6 15 0
7 15 0
8 15 0
9 15 0
10 15 0.01
11 15 0
12 15 0
13 15 0
14 15 0
15 15 0.98
16 15 0.01
17 15 0
18 15 0
19 15 0
0 16 0
1 16 0
2 16 0.02
3 16 0
4 16 0
5 16 0
6 16 0.01
7 16 0.01
8 16 0
9 16 0
10 16 0.05
11 16 0.01
12 16 0.01
13 16 0
14 16 0
15 16 0
16 16 0.89
17 16 0.01
18 16 0
19 16 0
0 17 0
1 17 0
2 17 0
3 17 0
4 17 0
5 17 0
6 17 0
7 17 0
8 17 0
9 17 0
10 17 0
11 17 0
12 17 0
13 17 0
14 17 0
15 17 0
16 17 0
17 17 1
18 17 0
19 17 0
0 18 0
1 18 0.01
2 18 0
3 18 0
4 18 0
5 18 0
6 18 0
7 18 0
8 18 0
9 18 0
10 18 0.01
11 18 0
12 18 0
13 18 0
14 18 0
15 18 0
16 18 0.01
17 18 0
18 18 0.98
19 18 0
0 19 0
1 19 0
2 19 0
3 19 0
4 19 0
5 19 0
6 19 0
7 19 0
8 19 0
9 19 0
10 19 0
11 19 0
12 19 0
13 19 0
14 19 0
15 19 0
16 19 0
17 19 0
18 19 0
19 19 1
         };
    \end{axis}
\end{tikzpicture}
%
    \caption{Confusion matrix for Random Forest ensemble} 
    \label{fig:ensemble_cm}
\end{figure}
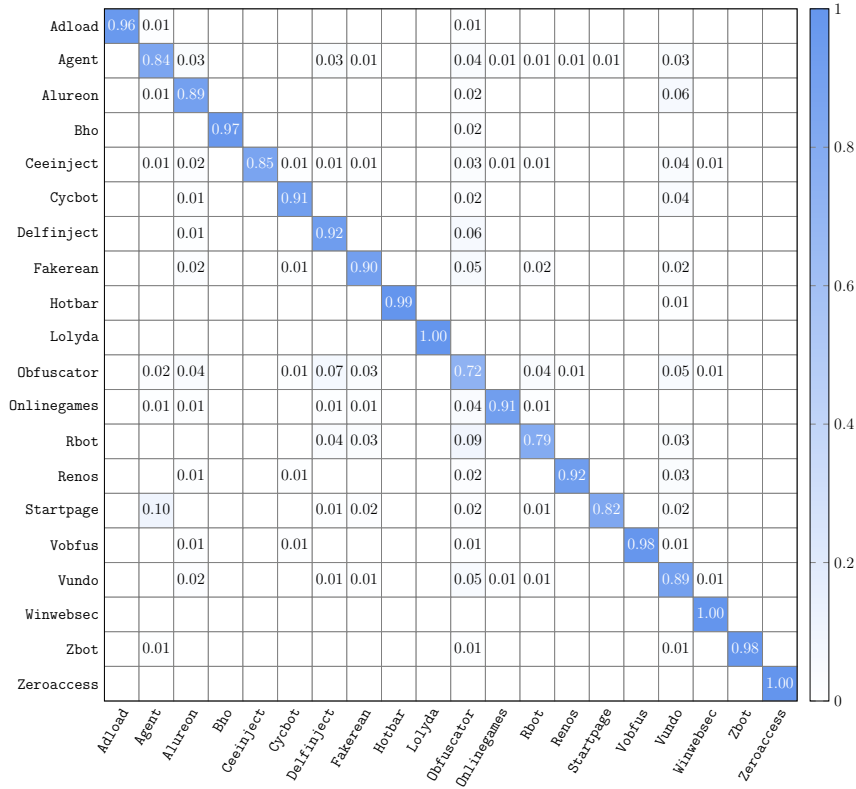

As a second ensemble experiment, we generate feature vectors using the~7 models
and train another classifier. AC-GAN and Resnet152 produce~$20\times1$ 
output vectors, while all other models generate one class label. 
These feature vectors are concatenated and used to 
train a Random Forest (RF) model. This RF based ensemble yields an accuracy of~92.09\%.
Interestingly, this RF model provides a noticeable improvement
on the more difficult families, although it performs somewhat worse
than Resnet152 on the easier families, including Cycbot, Adload, and Alureon.
We provide the confusion matrix for this model in Figure~\ref{fig:ensemble_cm}.
Apparently, the ``noise'' from the other other models causes the RF classifier to 
make more mistakes than Resnet152 on the easier models. 


\subsection{Generative Image Performance}\label{sect:GIP}

\begin{figure}[!htb]
    \centering
    \begin{tabular}{ccc}
        \includegraphics[width=0.2\textwidth]{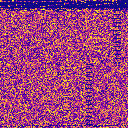}
        \includegraphics[width=0.2\textwidth]{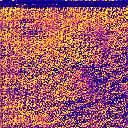}
& &
        \includegraphics[width=0.2\textwidth]{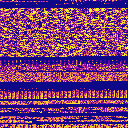}
        \includegraphics[width=0.2\textwidth]{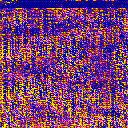}
\\
\\[-1.5ex]
    (a) Real vs generated image
& &
    (b) Real vs generated image
    \end{tabular}
    \caption{Real vs fake images (DC-GAN and Adload family)}\label{fig:dcgan_generated}
\end{figure}

Recall that DC-GAN is an unsupervised architecture, which implies that
we do not directly obtain accuracy scores 
from a DC-GAN discriminator model. However, 
we can use DC-GAN to analyze the generative power
of our AC-GAN generators. Figure~\ref{fig:dcgan_generated} shows a comparison between real and 
generated images using DC-GAN. 

To analyze the performance of our generator, below
we experiment with binary classification, i.e., malware vs benign. We use our dataset 
of~704 benign executables, extract images from them using the Colormap method then 
use SVM and CNN to classify them with our malware dataset. After that, we mix 
the benign samples with fake malware images from AC-GAN, DC-GAN and compare 
the detection performance. 

\begin{figure}[!htb]
    \centering
    \begin{tabular}{ccc}
        \includegraphics[width=0.2\textwidth]{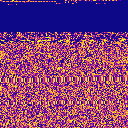}
        \includegraphics[width=0.2\textwidth]{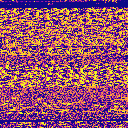}
& &
        \includegraphics[width=0.2\textwidth]{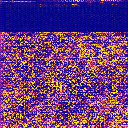}
        \includegraphics[width=0.2\textwidth]{images/generated_ceeinject_truncated1.png}
\\
\\[-1.5ex]
    (a) Real images
& &
    (b) Fake images
    \end{tabular}
    \caption{Real and fake images of Ceeinject family}
    \label{fig:ceeinject_truncated}
    \end{figure}

\begin{figure}[!htb]
    \centering
    \begin{tabular}{c}
        \includegraphics[width=0.14\textwidth]{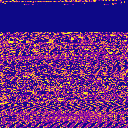}
        \includegraphics[width=0.14\textwidth]{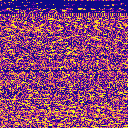}
        \includegraphics[width=0.14\textwidth]{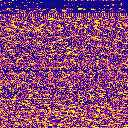}
\\
    (a) Real images
\\
\\[-1.5ex]
        \includegraphics[width=0.14\textwidth]{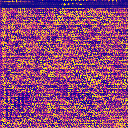}
        \includegraphics[width=0.14\textwidth]{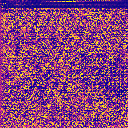}
        \includegraphics[width=0.14\textwidth]{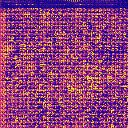}
\\
    (b) Fake images
    \end{tabular}
    \caption{Real and fake color images of Adload family}\label{fig:adload_truncated}
    \end{figure}

Recall that AC-GAN is a multiclass GAN, and hence it uses class labels when generating
images. Figure~\ref{fig:ceeinject_truncated} and Figure~\ref{fig:adload_truncated}
show real and AC-GAN generated images from the Ceeinject and Adload family,
respectively. 

\subsubsection{Binary Classification Results}

For our first set of binary classification experiments,
our dataset consists of~10,000 malware samples from~20 families (approximately~500 samples
per family) and~704 benign samples. 
We use colormap images and SVM for these malware detection experiment. 
We find that the 5-fold cross validation results
range from~99\% to~100\%, both in terms of accuracy and AUC. 

Next, we train models to distinguish between~10,000 fake malware images 
and~10,000 real malware images. Here, we ignore the benign set and
are trying to distinguish between real and fake images.
In this case, an SVM model still scores~99\%\ to~100\%\ in accuracy and AUC. 
This shows that GAN-generated images are surprisingly easy to distinguish from
real images.

We also tried to generate fake images using other variations of GANs, 
such as DC-GAN and Wasserstein GANs with gradient penalty (WGAN-GP). 
As WGAN-GP requires intensive computation power and 
training time, we only trained it for~100 epochs, with~300 epochs for DC-GAN. The results 
are still~99\%\ to~100\%\ for the accuracy and AUC. The colormap method 
may cause problems for the generated images because we only have~$16\times16$ available colors. 
However, when we experiment with the 3-gram method using AC-GAN, we obtain he similar results. 

\subsection{Discussion}

\begin{figure}[!htb]
    \centering
    \begin{tabular}{ccc}
        \multicolumn{3}{c}{\includegraphics[width=0.35\textwidth]{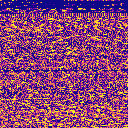}}
    \\
        \multicolumn{3}{c}{(a) Real image}
    \\
    \\[-1.125ex]
        \includegraphics[width=0.35\textwidth]{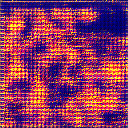}
    & &
        \includegraphics[width=0.35\textwidth]{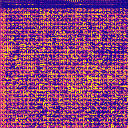}
    \\
    (b) Generated (30 epochs)
    & &
    (c) Generated (200 epochs)
    \end{tabular}
    \caption{Real and fake color images of Adload family}
\label{fig:epoch_comparison}
\end{figure}

Intuitively, since GANs include both a generator and a discriminator, they 
should improve both models, as compared to a more standard approach. 
In our experiments, AC-GAN discriminator performs best with~30 epochs for 
training while the AC-GAN generator requires~200 epochs or more for high-quality image generation. 
Figure~\ref{fig:epoch_comparison} 
shows the comparison between a real image and generated images with different numbers of
epochs. At~200 epochs, the AC-GAN generator clearly performs better. However, the AC-GAN 
discriminator shows clear signs of overfitting at such a high number of epochs,
with the accuracy decreasing to~65\%.

As with our malware classification experiments, 
Resnet152 outperforms GANs in terms of
accuracy and F1-score, and Resnet152 also has 
a substantially faster training time. This again shows the 
clear advantage of a pre-trained model. This indicates that
image-based malware analysis should be 
should focus more on pre-trained models, such as Resnet152 or VGG19. 
AC-GAN is competitive as a classifier. However, considering the long training time 
for AC-GAN, Resnet152 is the clear winner in our experiments.

\section{Conclusion and Future Works}\label{chap:future}

We experimented with four different ways of extracting images from malware executables:
grayscale, colormap, 3-gram, and PE. We found that based on the same amount of data, 
the colormap method improves the overall results for all models, with AC-GAN showing the most 
improvement for the malware families classification problem. XGBoost and RBM were faster to train 
as compared to SVM or GAN; however for XGBoost the memory requirement is high. 

Based on the confusion matrices, our models are not doing well with malware families that 
are more general or obfuscated. For future work, experiments with transformers or 
long short term memory (LSTM) would be interesting to consider, especially 
with respect to the more highly obfuscated malware families. The PE image 
method did not perform well in our experiments, but we believe that this image
generation method has considerable room for improvement. 

There are many GAN variants (StyleGAN, MalGAN, etc.)
that could be considered, especially with respect to 
the malware image generation problem. 
Designing a new GAN architecture or changing the AC-GAN 
loss function to speed up the training process of its generator would 
also be interesting topics to explore. 


\bibliographystyle{plain}
\bibliography{references.bib}

\section*{Appendix}

In this appendix, we provide details of the AC-GAN 
discriminator and generator architectures.
Note that these architectures are specified to the case of color
images of size~$128\times 128$. Other images and sizes are
straightforward modifications. 

\begin{figure}[!htb]
    \centering
     \includegraphics[width=0.65\textwidth]{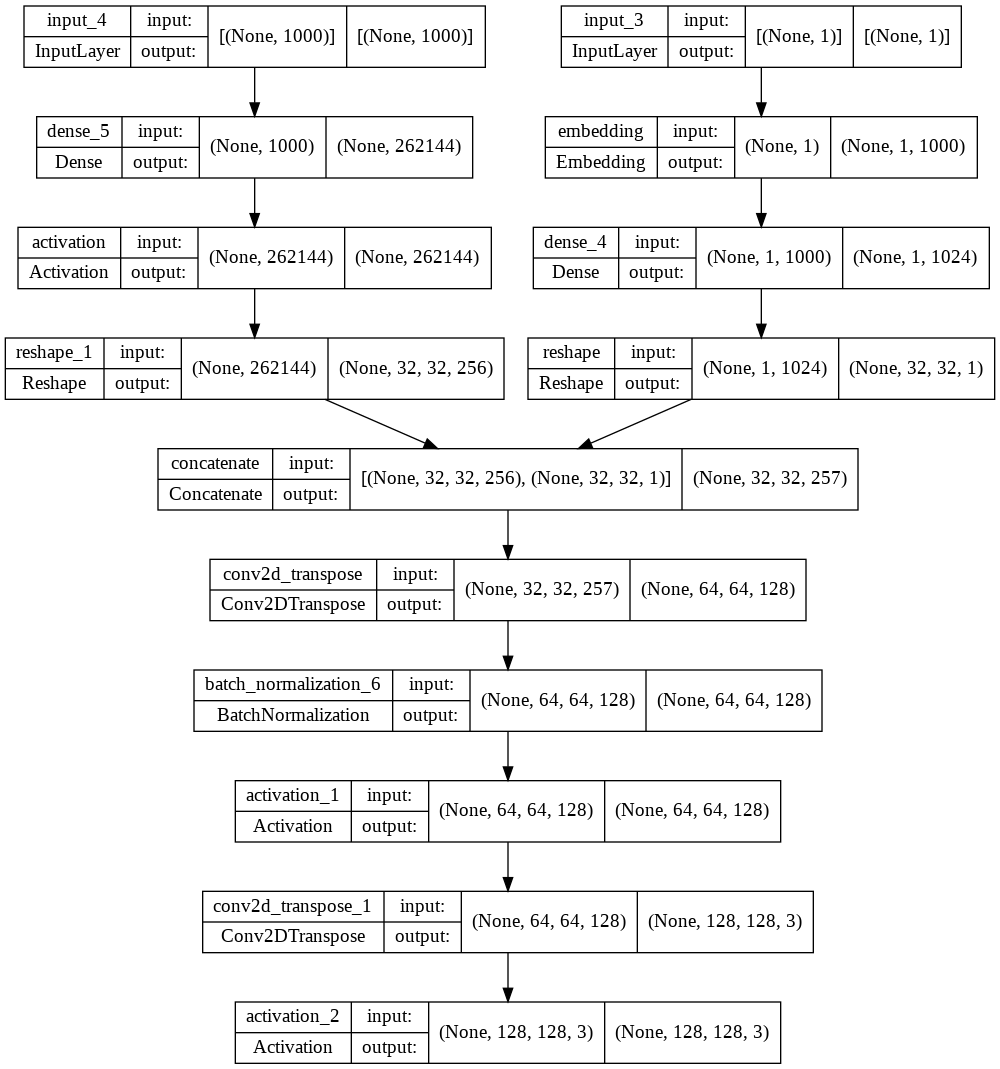}
\caption{AC-GAN generator architecture}
\label{fig:acgan_generator}
\end{figure}

\begin{figure}[!htb]
    \centering
     \includegraphics[width=0.575\textwidth]{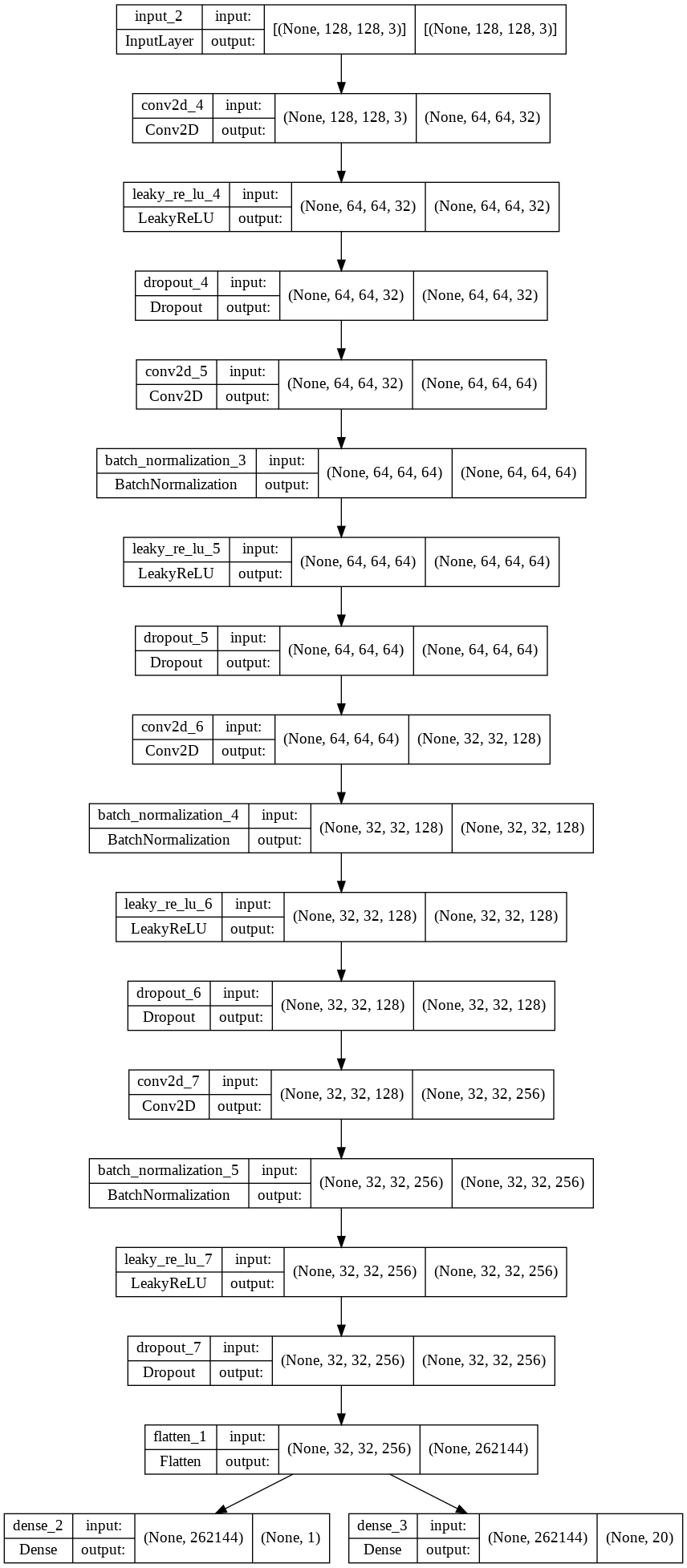}
\caption{AC-GAN discriminator architecture}
\label{fig:acgan_discriminator}
\end{figure}

\end{document}